\documentclass[reprint,superscriptaddress, amsmath,amssymb, aps]{revtex4-2}

\usepackage{xcolor}
\usepackage{listings}
\usepackage{amsmath}
\usepackage{mathtools}
\usepackage{graphicx}
\usepackage{dcolumn}
\usepackage{bm}
\usepackage{algorithm}
\usepackage{algpseudocode}
\usepackage{changepage}

\definecolor{codegreen}{rgb}{0,0.6,0}
\definecolor{codegray}{rgb}{0.5,0.5,0.5}
\definecolor{codepurple}{rgb}{0.58,0,0.82}
\definecolor{backcolour}{rgb}{0.95,0.95,0.92}
\newcommand{\ja}[1]{}

\lstdefinestyle{mystyle}{
 backgroundcolor=\color{backcolour}, 
 commentstyle=\color{codegreen},
 keywordstyle=\color{magenta},
 numberstyle=\tiny\color{codegray},
 stringstyle=\color{codepurple},
 basicstyle=\ttfamily\footnotesize,
 breakatwhitespace=false, 
 breaklines=true, 
 captionpos=b, 
 keepspaces=true, 
 numbers=left, 
 numbersep=5pt, 
 showspaces=false, 
 showstringspaces=false,
 showtabs=false, 
 tabsize=2
}

\begin{document}


\preprint{APS/123-QED}

\title{Biased versus unbiased numerical methods for stochastic simulations}

\author{Javier Aguilar}
\email{javieraguilar@ifisc.uib-csic.es}
\affiliation{Instituto de F\'{\i}sica Interdisciplinar y Sistemas Complejos IFISC (CSIC-UIB), Campus UIB, 07122 Palma de Mallorca, Spain. }
\affiliation{Investigador ForInDoc del Govern de les Illes Balears en el departamento de Electromagnetismo y Física de la Materia e Instituto Carlos I de Física Teórica y Computacional, Universidad de Granada, Granada E-18071, Spain .}
\author{Jos\'e J. Ramasco}
\affiliation{Instituto de F\'{\i}sica Interdisciplinar y Sistemas Complejos IFISC (CSIC-UIB), Campus UIB, 07122 Palma de Mallorca, Spain. }
\author{Ra\'ul Toral}
\affiliation{Instituto de F\'{\i}sica Interdisciplinar y Sistemas Complejos IFISC (CSIC-UIB), Campus UIB, 07122 Palma de Mallorca, Spain. }
\date{\today}

\begin{abstract}
\section*{Abstract}
Approximate numerical methods are one of the most used strategies to extract information from many-interacting-agents systems. In particular, {numerical approximations are} of extended use to deal with epidemic, ecological and biological models, since unbiased methods like the Gillespie algorithm can become unpractical due to high CPU time usage required. However, the use of approximations has been debated and there is no clear consensus about whether unbiased methods or biased approach is the best option. In this work, we derive scaling relations for the errors in approximations based on binomial extractions. This finding allows us to build rules to compute the optimal values of both the discretization time and number of realizations needed to compute averages with the {biased} method with a target precision and minimum CPU-time usage. Furthermore, we also present another rule to discern whether the unbiased method or {biased} approach is more efficient. Ultimately, we will show that the choice of the method should depend on the desired precision for the estimation of averages.
\end{abstract}

\maketitle

\section{Introduction}\label{sec:Intro}

Epidemic modeling has traditionally relied on stochastic methods to go beyond mean-field deterministic solutions \cite{britton2010stochastic,allen2008introduction,andersson2012stochastic,pastor2015epidemic,brauer2017mathematical}. The contagion process itself is naturally adapted to a stochastic treatment since the basic units, individuals, can not be described successfully using deterministic laws. For example, two given individuals may or may not develop a contact even though they are potentially able to do so given their geographical location. Even further, should the contact be established and should one of the individuals be infectious, the infection of the second individual is not a certainty, but rather an event that occurs with some probability. Computational epidiomiologists have implemented these stochastic contagions in all the modeling efforts and at  different scales, from agent-based \cite{eubank2004modelling,ferguson2005strategies,longini2005containing,germann2006mitigation, ciofi2008mitigation,merler2010role,aleta2020modelling} to population-based \cite{sattenspiel1995structured,colizza2006role,colizza2007,balcan2009multiscale,balcan2010}. In the case of agent-based models stochastic contagion events can be traced one by one, even though for practical purposes in computation sometimes they may be aggregated. In the population-level models, different contagion processes are aggregated together following some generic feature (number of neighboring individuals, geographical location, etc.). These models have the virtue of drastically reducing the number of variables needed to describe the whole population and, at the computational level, the positive side effect of enormously decreasing the model running time.  
A wide-spread practice nowadays~\cite{balcan2011phase,merler2015spatiotemporal,zhang2017spread,gomez2018critical,gilbert2020preparedness,chinazzi2020effect,arenas2020modeling,aguilar2022impact} is to approximate the statistical description of these contagion aggregations by binomial or multinomial distributions. 


The same methodological and conceptual issues appear well beyond the reign of epidemic modeling. Indeed, stochastic processes are one of the main pillars of complexity science~\cite{argun2021simulation,thurner2018introduction,tranquillo2019introduction}. The list of fruitful applications is endless, but just to name a few paradigmatic examples:  population dynamics in ecology~\cite{odum1971fundamentals,balaban2004bacterial}, gene expression~\cite{elowitz2002stochastic}, metabolism in cells~\cite{kiviet2014stochasticity}, finances and market crashes~\cite{rolski2009stochastic,sornette2003critical}, telecommunications~\cite{baccelli2010stochastic}, chemical reactions~\cite{van1992stochastic}, quantum physics~\cite{milz2021quantum} and active matter~\cite{ramaswamy2010mechanics}. As models become more intricate, there arises the technical challenge of producing stochastic trajectories in feasible computation times, since unbiased methods that generate statistically correct realizations of stochastic trajectories may become unpractical due to lengthy computations. Approximate methods aim at solving this issue by significantly reducing the CPU time usage. 
The use of approximated methods is extended (see e.g. \cite{gomez2018critical,balcan2009multiscale,balcan2010modeling}), and some authors assert that they might be the only way to treat effectively large systems of heterogeneous agents~\cite{gillespie2007stochastic}. However, other works claim that the systematic errors induced by the approximations might not trade-off the reduction in computation time~\cite{fennell2016limitations,gomez2011nonperturbative}.
The primary objective of this work is to shed light in this debate and assess in which circumstances {approximate methods based on binomial extractions, which we call \emph{binomial methods},} can be advantageous with respect to the unbiased algorithms. 

To solve this question, we derive in this paper a scaling relation for the errors of the binomial methods. This main result allows us to obtain optimal values for the discretization time and number of realizations to compute averages with a desired precision and minimum CPU time consumption. Furthermore, we derive a rule to discern if the binomial method is going to be faster than the unbiased counterparts. Lastly, we perform a numerical study to compare the performance of both the unbiased and binomial methods and check the applicability of our proposed rules. Ultimately, we will show that the efficiency of the binomial method is superior to the unbiased approaches only when the target precision is below a certain threshold value.

\subsection{Transition rates}\label{sec:transition_rates}

{Throughout this work we will focus on pure jumping processes, this is, stochastic models in which agents switch states within a discrete set of possible states. Spontaneous creation or annihilation of agents will not be considered, therefore, its total number, $N$, is conserved. We furthermore assume Markovian dynamics, so given that the system is in a particular state at time $t$, the ``microscopic rules" that dictate the switching between states just depend on the current state $\boldsymbol{s}(t)=\left\{s_1(t),\dots,s_N(t)\right\}$. These microscopic rules are given in terms of the transition rates, defined as the conditional probabilities per unit of time to observe a transition,

\begin{equation}
 \label{eq:def_transition_rate}
 w^t\left(\boldsymbol{s}\rightarrow \boldsymbol{s'}\right) := \lim_{dt\rightarrow 0} \frac{P( \boldsymbol{s'};t+dt| \boldsymbol{s};t)}{dt}.
\end{equation}

 A particular set of transitions in which we are specially interested define the ``one-step processes", meaning that the only transitions allowed are those involving the change of a single agent's state, with rates

\begin{align}
\label{eq:def_one_step_transition_rates}
 &w^t_i(s_i\to s_i') :=\nonumber\\ &w^t(\{s_1,\dots,s_i,\dots,s_N\}\to\{s_1,\dots,s_i',\dots,s_N\}),
\end{align}

for $i=1,\dots,N$. Our last premise is to consider only transition rates $w_i(s_i\to s_i') $ that do not depend explicitly on time $t$. Note that the rates could, in principle, be different for every agent and depend in an arbitrary way on the state of the system. The act of modelling is actually to postulate the functional form of these transition rates. This step is conceptually equivalent to the choice of a Hamiltonian in equilibrium statistical mechanics.}

{Jumping processes of two-state agents, such that the possible states of the $i^{th}$ agent can be $s_i=0$ or $s_i=1$, are widely used in many different applications, such as: protein activation~\cite{leff1995two}, spins $1/2$~\cite{brush1967history}, epidemic spreading~\cite{keeling2011modeling,pastor2015epidemic}, voting dynamics~\cite{fernandez2014voter}, chemical reactions~\cite{lee1996two,huang2000action}, drug-dependence in pharmacology~\cite{bridges2008g}, etc. For binary-state systems, quite commonly, the rate of the process $s_i=0\to s_i=1$ is different of the reverse process $s_i=1\to s_i=0$ and we define the rate of agent $i$ as}

\begin{align}
 w_i(s_i)&:= \begin{cases}
 w_i(0\to 1)\quad \text{ if } s_i=0,\\
 w_i(1\to 0)\quad \text{ if } s_i=1.
 \end{cases}
\end{align}

As a detailed observation is usually unfeasible, we might be interested on a macroscopic level of description focusing, for example, on the occupation number $n(t)$, defined as the total number of agents in state $1$,

\begin{equation}
 \label{eq:def_n}
 n(t) := \sum_{i=1}^N s_i(t),
\end{equation}

being $N-n(t)$ the equivalent occupation of state $0$. In homogeneous systems, those in which $w_i(s_i)=w(s_i),\, \forall i$, transition rates at this coarser level can be computed from those at the agent-level as

\begin{align}
\label{eq:def_macroscopic_transition_rates}
 W(n\rightarrow n+1) &= (N-n) w(0), \nonumber \\
 W(n\rightarrow n-1) &= n w(1).
\end{align}

Some applications might require an intermediate level of description between the fully heterogeneous [Eq.~\ref{eq:def_one_step_transition_rates}] and the fully homogeneous [Eq.~\eqref{eq:def_macroscopic_transition_rates}]. In order to deal with a coarse-grained heterogeneity, we define $\mathcal{C}$ different classes of agents. Agents can be labeled in order to identify their class, so that $l_i=\ell$ means that the $i^{th}$ agent belongs to the class labeled $\ell$ with $\ell\in[1,\mathcal{C}]$ and we require that all agents in the same class share the same transition rates $w_i(s_i)=w_\ell(s_i),\, \forall l_i=\ell$. This classification allows us to define the occupation numbers $N_\ell$ and $n_\ell$ as the total number of agents of the $\ell^{th}$ class and the number of those in state $1$ respectively. Moreover, we can write the class-level rates:

\begin{align}
\label{eq:def_class_transition_rates_n}
 W_{\ell}(n_\ell\rightarrow n_\ell+1) &= \left(N_\ell-n_\ell\right) w_{\ell}(0), \nonumber \\
 W_{\ell}(n_\ell\rightarrow n_\ell-1) &= n_\ell w_{\ell}(1).
\end{align}

In general, stochastic models are very difficult and can not be solved analytically. Hence, one needs to resort to numerical simulations than can provide suitable {estimations} to the quantities of interest. There are two main types of simulation strategies: unbiased continuous-time and discrete-time algorithms. Each one comes with its own advantages and disadvantages that we summarize in the next sections. 

\section{Methods}
\subsection{Unbiased continuous-time algorithms}\label{sec:continuous_time_algorithms}
We proceed to summarize the main ideas behind the unbiased continuous-time algorithms, and refer the reader to ~\cite{gillespie1976general,gillespie2007stochastic,toral2014stochastic,keeling2011modeling,cota2017optimized,colizza2008epidemic,tailleur2009simulation,masuda2022gillespie} for a detailed description. Say that we know the state of the system $\boldsymbol{s}(t)$ at a given time $t$. Such state will remain unchanged until a random time ${t'}>t$, when the system experiences a transition or ``jump" to a new state, also random, $\boldsymbol{s}'(t')$: 

\begin{equation}
 \boldsymbol{s}(t)\xrightarrow{{t'}-t}\boldsymbol{s'}({t'}).
\end{equation}

Therefore, the characterization of a change in the system necessarily requires us to sample both the transition time $\Delta {t} = {t'}-t$ and the new state $\boldsymbol{s'}({t'})$. 

For binary one-step processes, new states are generated by changes in single agents $s_i\to 1-s_i$. The probability that agent $i$ changes its state in a time interval ${t'}\in [t,t+dt]$ is $w_i\left(s_i\right) dt$ by definition of transition rate. Therefore, the probability that the agent will not experience such transition in an infinitesimal time interval is $1-w_i\left(s_i\right) dt$. Concatenating such infinitesimal probabilities, we can compute the probability $Q_i(s_i,\Delta t)$ that a given agent does not change its state during an arbitrary time lapse $\Delta t$ as well as the complementary probability $P_i(s_i,\Delta t)$ that it does change state as

\begin{align}
 \label{eq:prob_remaining_constant}
 & Q_i(s_i,\Delta t) =\lim_{dt\rightarrow0} (1-w_i(s_i)dt)^{\Delta t/dt}=e^{-w_i(s_i)\Delta t},\nonumber\\
 &P_i(s_i,\Delta t)=1-e^{-w_i(s_i)\Delta t}.
\end{align}

Eq.~(\ref{eq:prob_remaining_constant}) conforms the basic reasoning from which most of the continuous-time algorithms to simulate stochastic trajectories are built. It allows us to extend our basic postulate from Eq.~(\ref{eq:def_transition_rate}), which only builds probabilities for infinitesimal times ($dt$), to probabilities of events of arbitrary duration ($\Delta t$). It is important to remark that Eq.~(\ref{eq:prob_remaining_constant}) is actually a conditional probability: it is only valid provided that there are no other updates of the system in the interval $\Delta t$. From it we can also compute the probability density function that the $i^{th}$ agent remains at $s_i$ for a non-infinitesimal time $\Delta t$ and then experiences a transition to $s_i'=1-s_i$ in the time interval $[t+\Delta t,t+\Delta t+dt]$:

\begin{align}
 f_i(s_i;\Delta t) = e^{-w_i(s_i)\Delta t}w_i(s_i).
\end{align}

The above quantity is also called first passage distribution for the $i^{th}$ agent. Therefore, given that the system is in state $\boldsymbol{s}$ at time $t$, one can use the elements defined above to compute the probability that the next change of the system is due to a switching in the agent $i$ at time $t'\in [t+\Delta t,t+\Delta t+dt]$:

\begin{align} \label{eq:distribution_first_jump}
 & \text{P($i^{th}$ agent switches state in $[t+\Delta t,t+\Delta t+ dt]$) }\times \nonumber \\ 
 & \text{P(Other agents change state only after $t+\Delta t+dt$) }= \nonumber \\
 & f_i(s_i;\Delta t) dt \times \prod_{j\ne i}^N Q_j(s_j,\Delta t)= e^{-W (\boldsymbol{s})\Delta t}w_i(s_i) dt,
\end{align}

where we have defined the total exit rate,

\begin{equation}\label{eq:exit_rate}
W(\boldsymbol{s}) := \sum_{i=1}^N w_i(s_i).
\end{equation}

Two methods, namely the first-reaction method and {the Gillespie algorithm}, can be distinguished based on the scheme used to sample the random jumping time ${t'}$ and switching agent $i$ from the distribution specified in Eq.~(\ref{eq:distribution_first_jump}). The first-reaction method involves sampling one tentative random time per transition {using $f_i(s_i;\Delta t)$} and then choosing the minimum among them as the transition time {and reaction} that actually occurs. In contrast, the Gillespie algorithm directly samples the transition time {using the total rate $W(\boldsymbol{s})$} and then determines which transition is being activated. {Depending on the algorithm used to randomly select the next reaction, the computational complexity of the \ja{unbiased methods can vary from linear to constant} in the number of reactions (see e.g.~\cite{masuda2022gillespie}). Through the rest of the manuscript, we will use Gillespie algorithms with binary search in representation of unbiased methods.}

\subsection{Discrete-time approximations}\label{sec:discrete_time_algorithms}
In this section, we consider algorithms which at simulation step $j$ update time by a constant amount, $t_{j+1}=t_{j}+\Delta t$. Note that the discretization step $\Delta t$ is no longer stochastic, and it has to be considered as a new parameter that we are in principle free to choose. Larger values of $\Delta t$ result in faster simulations since fewer steps are needed in order to access enquired times. Nevertheless, the discrete-time algorithms introduce systematic errors that grow with $\Delta t$. 

\subsubsection{Discrete-synchronous}
It is possible to use synchronous versions of the process where all agents can potentially update their state at the same time $t_j$ using the probabilities $P_i(s_i,\Delta t)$ defined in Eq.~\eqref{eq:prob_remaining_constant} (see e.g. \cite{colizza2008epidemic,colizza2007reaction,goutsias2013markovian}). 
\begin{algorithm}[H]
 \caption{Discrete time synchronous agent level}
 \label{al:DSA}
 \begin{algorithmic}[1]
 \item Increment time: $t_{j+1}=t_j+\Delta t$
 \item Compute all probabilities $P_i(s_i,\Delta t)$, $i=1,\dots,N$, using Eq.~\eqref{eq:prob_remaining_constant}.
 \item For all agents, generate a uniform random number $\hat{u}_i\in[0,1]$. If $\hat{u}_i<P_i(s_i,\Delta t)$ change the state $s_i\to 1-s_i$.
 \item go to 1.
 \end{algorithmic}
\end{algorithm}

We note that the use of synchronous updates changes the nature of the process since simultaneous updates were not allowed in the original continuous-time algorithms. Given that the probabilities $P_i(s_i,\Delta t)$ tend to zero as $\Delta t\to 0$, one expects to recover the results of the continuous-time asynchronous approach in the limit $\Delta t\rightarrow 0$. Nevertheless, users of this method should bear in mind that this approximation could induce discrepancies with the continuous-time process that go beyond statistical errors~\cite{allen1994some}.

\subsubsection{Binomial method: two simple examples}
When building the class-version of the synchronous agent level (Algorithm~\ref{al:DSA}), one can merge together events with the same transition probability and sample the updates using binomial distributions. This is the basic idea behind the binomial method, which is of extended use in the current literature (e.g.~{\cite{aguilar2022impact,balcan2010modeling,chinazzi2020effect,leier2008generalized,peng2007efficient}}). Since references presenting this method are scarce, we devote a longer section to its explanation.

Let us start with a simple example. Say that we are interested in simulating the decay of $N$ radioactive nuclei. We denote by $s_i=1$ that nucleus $i$ is non-disintegrated and by $s_i=0$ the disintegrated state. All nuclei have the same time-independent decay rate $\mu$:

\begin{align}\label{eq:decay}
 w_i(1\rightarrow 0) = \mu, \quad w_i(0\rightarrow 1) = 0.
\end{align}

This is, all nuclei can decay with the same probability $\mu dt$ in every time-bin of infinitesimal duration $dt$, but the reverse reaction is not allowed. This simple stochastic process leads to an exponential decay of the average number $n_t$ of active nuclei at time $t$ as $\langle n_t\rangle =Ne^{-\mu t}$.

Using the rates \eqref{eq:decay}, we can compute the probability that one nucleus disintegrates in a non-infinitesimal time $\Delta t$ [Eq.~\ref{eq:prob_remaining_constant}],

\begin{equation}
 \label{eq:P_Having_a_desintegration_inn_Dt}
 p:=P_i(1,\Delta t)= 1 - e^{-\mu \Delta t}, \quad \forall i.
\end{equation}

Therefore every particle follows a Bernoulli process in the time interval $\Delta t$. \ja{That is,} each particle decays with a probability $p$ and remains in the same state with a probability $1-p$. {As individual decays are independent of each other, }the total number of decays in a temporal-bin of duration $\Delta t$ follows a binomial distribution ${\bf B}(N,p)$,

\begin{equation}\label{eq:binomial}
 P[n \text{ decays in $\Delta t$}]={N\choose n}p^n(1-p)^{N-n}.
\end{equation}

The average of the binomial distribution is $\langle n\rangle =Np$ and its variance $\sigma^2[n]=Np(1-p)$.
This result invites to draw stochastic trajectories with a recursive relation: 

\begin{equation}
\label{eq:radioactive_decay_with_binomials}
 n_{t+\Delta t}=n_t-\Delta n_t,
\end{equation}

where we denote by $\Delta n_t\sim {\bf B}(n_t, p)$ a random value drawn from the binomial distribution, with average value $\langle \Delta n_t\rangle=n_t p$, and we start from $n_0=N$. In this simple example, it turns out that Eq.~(\ref{eq:radioactive_decay_with_binomials}) does generate unbiased realizations of the stochastic process. From this equation we obtain

\begin{equation}
 \langle n_{t+\Delta t} \rangle_B=\langle n_{t}\rangle_B-\langle \Delta n_{t}\rangle_B=\langle n_{t}\rangle_B(1-p).
\end{equation}

\begin{figure*}
 \centering\includegraphics[scale=1]{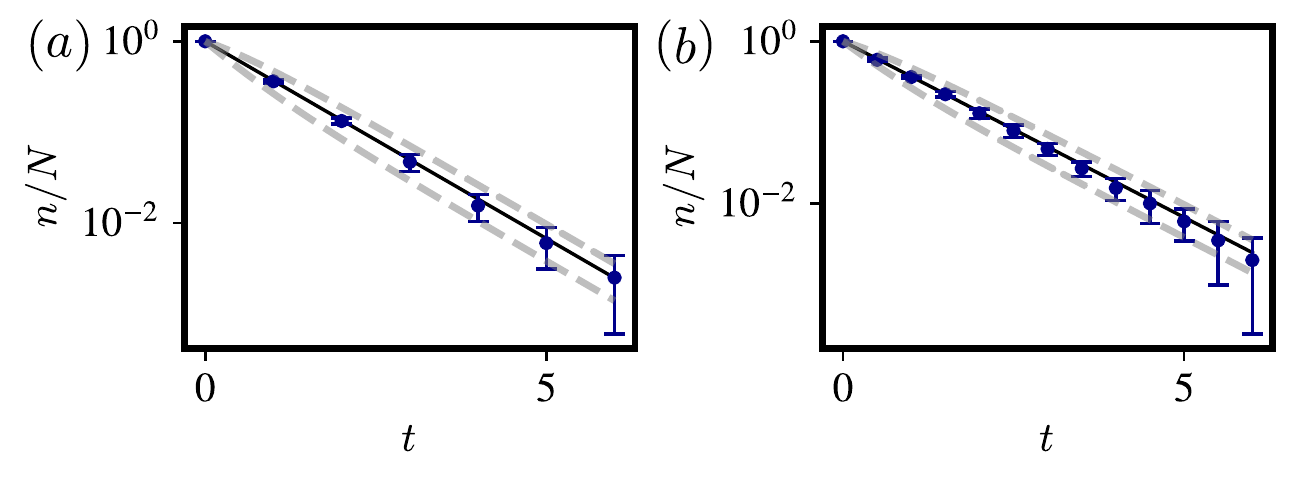}
 \caption{\textbf{Exact binomial method.} Simulations of the radioactive decay process with rates given by Eq.\eqref{eq:decay}, using the binomial method [Eq.~\eqref{eq:radioactive_decay_with_binomials}]. In (a) the time discretization is $\Delta t=1$, whereas in (b) is $\Delta t=0.5$. In both panels $N=100$ and $\mu=1$. Dots and error bars indicate the average and standard error respectively, both computed from 20 simulations. With {continuous} line, we show the analytical average (black) plus and minus the analytical standard error (gray {dashed lines}): $\langle n(t)\rangle\pm\sigma[n(t)]/\sqrt{20}$. Independently of the discretization time, the results from simulations agree with the analytical value within errors.}
 \label{fig:radioactive_decay}
\end{figure*}

The symbol $\langle \cdot \rangle_B$ \ja{denotes} averages over the binomial method. The solution of this recursion relation with initial condition $n_0=N$ is

\begin{equation}\label{eq:MF_radioactive_decay}
 \langle n_{t} \rangle_B=N\left(1-p\right)^\frac{t}{\Delta t}=N e^{-\mu t},
\end{equation}

which coincides with the exact result independently of the value of $\Delta t$.
Therefore, the choice of $\Delta t$ is just related to the desired time resolution of the trajectories. If $\Delta t \ll (N \mu)^{-1}$, many of the outcomes $\Delta n_t$ used in Eq.~(\ref{eq:radioactive_decay_with_binomials}) will equal zero as the resolution would be much smaller than the mean time between disintegration events. Contrary, if $\Delta t \gg (N\mu)^{-1}$, much of the information about the transitions will be lost and we would generate a trajectory with abrupt transitions. Still, both simulations would faithfully inform about the state of the system at the enquired times [see Figs.~\ref{fig:radioactive_decay} (a) and (b)].

Let us now apply this method to another process where it will no longer be exact. Nevertheless, the basic idea of the algorithm is the same: compute non-infinitesimal increments of stochastic trajectories using binomial distributions. We consider a system with $N$ agents which can jump between states with homogeneous constant rates:

\begin{align}\label{eq:birthdeath}
 w_i(1\to 0) = \mu,\quad w_i(0\to 1) = \kappa,
\end{align}

{Which, at macroscopic level read

\begin{align}
    W(n\to n-1) = n\mu,\quad W(n\to n+1) = (N-n)\kappa,
\end{align}
}
\ja{from which we can see that this process is a birth-death process. }Reasoning as before, the probabilities that a particle changes state in a non-infinitesimal time $\Delta t$ are:

\begin{align} \label{eq:two_states_transition_probabilities}
 P(0,\Delta t)= 1-e^{-\kappa \Delta t}, \nonumber \\
 P(1,\Delta t)= 1-e^{-\mu \Delta t}.
\end{align}

Where we can avoid the use of subscripts since all agents share the transition rates. At this point, we might feel also invited to write an equation for the evolution of agents in state $1$ in terms of the stochastic number of transitions:

\begin{equation}
 \label{eq:two_states_with_binomials}
 n_{t+\Delta t}=n_t+\Delta n_{t,0}-\Delta n_{t,1}.
\end{equation}

\begin{figure}[h]
 \centering
 \includegraphics[scale=1]{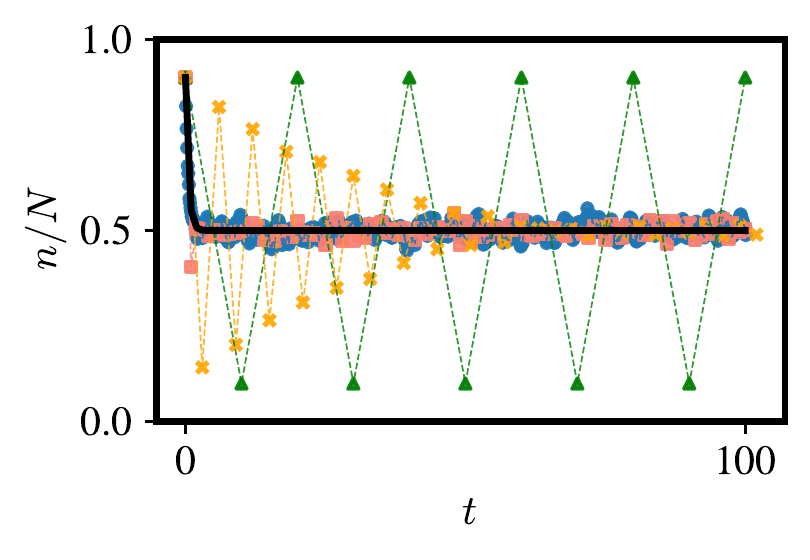}
 \caption{\textbf{Biased binomial method.} {Four} realizations of the birth and death process with constant rates defined by Eq.~\eqref{eq:birthdeath} simulated with the use of the binomial method [Eq.~\eqref{eq:two_states_with_binomials}]. In this case, we also use different time discretizations $\Delta t$, and fixed $N=1000$, $\mu=1$, and $\kappa=1$. Note the numerical instability that shows up as wild oscillations in the numerical trajectories for large time steps $\Delta t=10$ (triangles){, and $\Delta t=3$ (crosses)}. Otherwise, there is a good agreement between simulations and the expected average value ({continuous black line}) for both $\Delta t=0.1,1$ (circles and squares respectively)}
 \label{fig:binomial_example}
\end{figure}

Where $\Delta n_{t,0}$ and $\Delta n_{t,1}$ are binomial random variables distributed according to ${\bf B}(N-n_t,P(0,\Delta t))$ and ${\bf B}(n_t,P(1,\Delta t))$, respectively. However, trajectories generated with Eq.~(\ref{eq:two_states_with_binomials}) turn out to be only an approximation to the original process. The reason is that the probability that a given number of transitions $0\rightarrow 1$ happen in a time window is modified as soon as a transition $1\rightarrow 0$ occurs (and vice-versa). If we now take averages in Eq.~(\ref{eq:two_states_with_binomials}), use the known averages of the binomial distribution and solve the resulting linear iteration relation for $\langle n_t\rangle_B$, we obtain:

\begin{equation} \label{eq:two_states_average}
 \langle n_t \rangle_B=\left(n_0-\frac{b}{a}\right)(1-a)^{t/\Delta t}+\frac{b}{a}
\end{equation}

with $a=2-e^{-\mu \Delta t}-e^{-\kappa \Delta t}$ and $b=N(1-e^{-\kappa \Delta t})$. It is true that in the limit $\Delta t\to 0$, this solution recovers the exact solution for the evolution equation of the average number \ja{of agents in state 1} for the continuous-time process, namely

\begin{align}\label{eq:MF_two_states}
\frac{d\langle n_t\rangle}{dt}&=-\mu\langle n_t\rangle +\kappa(N-\langle n_t\rangle),\nonumber \\
 \langle n_t\rangle&=\left(n_0-N\frac{\kappa}{\kappa+\mu}\right)e^{-(\kappa+\mu)t}+N\frac{\kappa}{\kappa+\mu},
\end{align}

but the accuracy of the discrete approximation depends crucially on the value of $\Delta t$. If, for instance, we take $\Delta t\gg \max(\kappa^{-1},\mu^{-1})$, then we can approximate $a\approx 2$, $b\approx N$, such that Eq.~\eqref{eq:two_states_average} yields

\begin{equation}
 \langle n_t \rangle_B= 
 \begin{cases}
 N-n_0, & \text{if}\ t/\Delta t \text{ odd}, \\
 n_0, & \text{if}\ t/\Delta t \text{ even},
 \end{cases}
\end{equation}

a numerical instability that shows up as a wild oscillation, see Fig.~\ref{fig:binomial_example}. 

Therefore, the fact that agents are independent and rates are constant is not sufficient condition to guarantee that the binomial method generates unbiased trajectories for arbitrary values of the discretization step $\Delta t$. Nevertheless, it is remarkable that the only condition needed to ensure that Eq.~(\ref{eq:two_states_with_binomials}) is a good approximation to the exact dynamics, Eq.~(\ref{eq:MF_two_states}), is that $\Delta t \ll \min(\kappa^{-1},\mu^{-1})$. Given than the system size $N$ does not appear in this condition, we expect the binomial method to be very efficient to simulate this kind of process if we take a sufficiently small value for $\Delta t$, independently \ja{of} the number of agents, see Fig.~\ref{fig:binomial_example}, where both $\Delta t=0.1,1$ produce a good agreement for $\mu=\kappa=1$.
By comparing the average value of the binomial method, Eq.\eqref{eq:two_states_average} with the exact value, Eq.\eqref{eq:MF_two_states}, we note that the error of the binomial approximation can be expanded in a Taylor series 
\begin{equation}
 \langle n_{t} \rangle_B-\langle n_{t} \rangle= \lambda \Delta t+\mathcal{O}(\Delta t^2).
\end{equation}

where the coefficient of the linear term $\lambda$ depends on $t$ and $N$, as well as on other parameters of the model. We will check throughout this work that a similar expansion of the errors in the binomial method holds for the case of more complex models.

\subsubsection{Binomial method: general algorithm}
If we go back to the general two-state process in which the functional form of the rates can have an arbitrary dependence on the state of the system, we can approximate the probability that the state of agent $i$ changes in a time interval $\Delta t$ by $P_i(s_i,\Delta t)$ [Eq.~\eqref{eq:prob_remaining_constant}]. If all these probabilities are different, we cannot group them in order to conform binomial samples. If, on the other hand, we can identify large enough classes $\ell=1,2,\dots,\mathcal{C}$ such that all {$N_\ell$} agents in the same class $\ell$ have the same rates $w_\ell(\boldsymbol{s})$, we can approximate the variation of the occupation number $n_\ell$ of each class during the time $\Delta t$ as the difference $\Delta n_{\ell, 0}-\Delta n_{\ell, 1}$ where $\Delta n_{\ell, 0}$ and $\Delta n_{\ell, 1}$ follow, respectively, binomial distributions $B(N_\ell-n_\ell,P_{\ell}(0,\Delta t))$ and $B(n_\ell,P_{\ell}(1,\Delta t))$, with $P_\ell(s_i,\Delta t)$ given by Eq.~\eqref{eq:prob_remaining_constant} using any agent $i$ belonging to class $\ell$. All class occupation numbers are updated at the same time step $j$, yielding the synchronous binomial algorithm, which reads:
\begin{algorithm}[H]
 \caption{Binomial synchronous class level}
 \label{al:BM}
 \begin{algorithmic}[1]
 \item Update time as $t_{j+1}=t_j+\Delta t$.
 \item[] For every class $\ell\in[1,\dots,\mathcal{C}]$:
 \item \hspace{10pt}Update the values of $P_{\ell}(1,\Delta t), P_{\ell}(0,\Delta t)$, using Eq.~\eqref{eq:prob_remaining_constant}.
 \item \begin{adjustwidth}{10pt}{}
 Update the number of agents as $n_\ell\to n_\ell+\Delta n_{\ell, 0}-\Delta n_{\ell, 1}$, where $\Delta n_{\ell,0}$ and $\Delta n_{\ell,1}$ are values of binomial random variables distributed according to ${\bf B}(N_\ell-n_\ell,P_{\ell}(0,\Delta t))$ and ${\bf B}(n_\ell,P_{\ell}(1,\Delta t))$, respectively.
 \end{adjustwidth}
\item go to 1.
 \end{algorithmic}
\end{algorithm}

A similar reasoning can be built departing from the knowledge that the number $n$ of occurrences of continuous-time independent processes {with constant rates} follows a Poisson distribution~\cite{van1992stochastic}, {namely $e^{-\Lambda}{\Lambda^n}/{n!}$, being the parameter $\Lambda$ of the Poisson distribution equal to the product of the rate times the time interval considered. Therefore, the number of firings of each class in the time interval $\Delta t$, $\Delta n_{\ell,1}$ and $\Delta n_{\ell,0}$, can be approximated by Poisson random variables with parameters $W_\ell(n_\ell\to n_\ell-1)\Delta t$ and $W_\ell(n_\ell\to n_\ell+1)\Delta t$, respectively.} This conception gives rise to the {$\tau$}-leaping algorithm~{\cite{gillespie2007stochastic,keeling2011modeling,cao2006efficient,cao2009discrete,lecca2013stochastic,rathinam2005consistency,goutsias2013markovian}} used in the context of chemical modeling. {Given that Poisson random variables are unbounded from above, the $\tau$-leaping algorithm may yield negative values for the occupation numbers $n_\ell$ (see e.g.~\cite{gillespie2007stochastic,goutsias2013markovian}). Consequently, our focus will be on the binomial method, which does not exhibit this drawback.} 

\section{Results and Discussion}
\subsection{The $\dfrac{27}{4}$ rule}\label{sec:27_4_rule}
The major drawback of the binomial method to simulate trajectories is the necessity of finding a proper discretization time $\Delta t$ that avoids both slow and inaccurate implementations. In this section, we propose a semi-empirical predictor for the values of the optimal choice of $\Delta t$ that propitiates the smallest computation time for a fixed desired accuracy. Moreover, we will present a rule to discern whether an unbiased continuous-time algorithm or the discrete-time binomial method is more suitable for the required task.

Consider that we are interested in computing the average value $\langle Z\rangle$ of a random variable $Z$ that depends on the stochastic trajectory in a time interval $[0,T]$. For example, $Z$ could be the number of nuclei for the process defined in Eq.~\eqref{eq:decay} at a particular time $t\in[0,T]$. {We remark that $\langle Z\rangle$ could also stand for the central second moment of some random variable, thus accounting for fluctuations around some mean value. Also, the average $\langle Z\rangle$ could represent probabilities if $Z$ is chosen to be an indicator function (see e.g.~\cite{asmussen2007stochastic}).}

The standard approach to compute $\langle Z\rangle$ numerically generates $M$ independent realizations of the stochastic trajectories and measures the random variable $Z^{(i)}$ in each trajectory $i=1,\dots,M$. The average value $\langle Z\rangle$ is then approximated by the sample mean 

\begin{equation}\label{eq:averageZ_linear}
 Z_M:= \frac{1}{M}\sum_{i=1}^{M}Z^{(i)}.
\end{equation}

Note that $Z_M$ itself should be considered a random variable as its value changes from a set of $M$ realizations to another. 

For an unbiased method, such as Gillespie, the only error $\varepsilon$ in the estimation of $\langle Z\rangle$ by $Z_M$ is of statistical nature and can be computed from the standard deviation of $Z_M$, namely

\begin{equation}\label{eq:def_precision}
 \varepsilon=\frac{\sigma}{\sqrt{M}},\text{ with }
 \sigma:=\sqrt{\langle Z^2\rangle-\langle Z\rangle^2}.
\end{equation}

The quantification of the importance of the error, for sufficiently large $M$, follows from the central limit theorem~\cite{asmussen2007stochastic,toral2014stochastic} using the confidence intervals of a normal distribution:

\begin{equation}\label{eq:confidence_interval}
 P\left[\langle Z \rangle-\varepsilon\le Z_M \le \langle Z \rangle+\varepsilon\right] = 0.6827\dots 
\end{equation}

It is in this sense, that one says that the standard error $\varepsilon$ is the precision of the estimation and writes accordingly

\begin{equation}
 \langle Z \rangle=Z_M\pm \varepsilon. 
\end{equation}

Note that, according to Eq.\eqref{eq:def_precision}, for an unbiased method the error in the estimation of the sample mean $Z_M$ tends to zero in the limit $M\to\infty$.

For a biased method, such as the binomial, that uses a finite discretization time $\Delta t$ and generates $M_B$ independent trajectories, the precision is altered by a factor that does not tend to zero in the limit $M_B\to\infty$. Based on the result found in the simple birth and death example of the previous section, let us assume for now that this factor scales linearly with the discretization time $\Delta t$ and can be written as $\lambda\Delta t$ where $\lambda$ is a constant depending on the model. We will corroborate this linear assumption for the binomial method both with calculations and numerical simulations in the next section, \ja{and we refer to Supplementary Note 1 for a more general discussion in the case of a method with a possible non-linear dependence.} Then we can write the estimator using the binomial method as
{

\begin{equation}\label{eq:average_biased}
 \langle Z \rangle=Z_{M_B}+\lambda \Delta t\pm\varepsilon_{B},
\end{equation}

}
where $\varepsilon_{B}=\dfrac{\sigma}{\sqrt{M_B}}$ and $Z_{M_B}$ is the sample average, Eq.~\eqref{eq:averageZ_linear}, using $M_B$ realizations. The maximum absolute error of the biased method is then $|\lambda|\Delta t+\varepsilon_{B}$. Due to the presence of a bias term in the error, the only way that the precision of the binomial method can equal the one of an unbiased approach is by increasing the number of realizations $M_{B}$ compared to the number of realizations $M$ of the unbiased method. Matching the values of the errors of the unbiased and the biased methods, we arrive at the condition that the required number of steps of the biased method is

\begin{equation}\label{eq:number_of_realizations} M_B=M\left(\frac{|\lambda|\Delta t}{\varepsilon}-1\right)^{-2},
\end{equation}

and the additional requirement {$\Delta t<\frac{\varepsilon}{|\lambda|}$} (otherwise the bias is so large that it can not be compensated by the increase in the number of realizations $M_B$).

What a practitioner needs {is to compare the CPU times that the biased and unbiased methods require to achieve the same accuracy $\varepsilon$. For the biased method with a fixed time step $\Delta t$, the CPU time $t_B^{\text{(CPU)}}$} needed to generate one stochastic trajectory is proportional to the number of steps, $\dfrac{T}{\Delta t}$, needed to reach the final time $T$ and can be written as $C_B\dfrac{T}{\Delta t}$, where $C_B$ is the CPU time needed to execute one iteration of the binomial method. Hence the total time required to generate $M_B$ trajectories is
\begin{equation}\label{eq:cpu_t_binomial}
 t_B^{\text{(CPU)}}=C_B M_B \frac{ T}{\Delta t}.
\end{equation}
\ja{(Note that in massive parallel architectures, it might be possible to obtain a sub-linear dependence of the total time with the number of realizations $M$. This possibility is discussed in Supplementary Note 1.)}

The discretization time associated with a minimum value of the CPU time consumption and subject to the constraint of fixed precision is obtained by inserting Eq.~\eqref{eq:number_of_realizations} in Eq.~\eqref{eq:cpu_t_binomial} and minimizing for $\Delta t$ (see Supplementary Note 1). The optimal time reads:

\begin{equation}\label{eq:optimal_Dt}
 \Delta t^{\text{opt}}=\frac{1}{3} \frac{\varepsilon}{|\lambda|}.
\end{equation}

Inserting the equation for the optimal $\Delta t$ in Eq.~\eqref{eq:number_of_realizations}, one obtains:

\begin{equation}\label{eq:number_of_realizations_optimal_dt}
 M_B^{\text{opt}}=\frac{9}{4} M=\frac{9}{4}\left(\frac{\sigma}{\varepsilon}\right)^2,
\end{equation}

{which, remarkably, does not depend of $\lambda$ or other parameters.} Eqs.~\eqref{eq:optimal_Dt} and~\eqref{eq:number_of_realizations_optimal_dt} have major practical use, since they tell us how to choose $\Delta t^{\text{opt}}$ and $M_B^{\text{opt}}$ to use the binomial method to reach the desired precision $\varepsilon$ and with minimum CPU time usage.

Still, one important question remains. Provided that we use the optimal pair ($M_B^{\text{opt}}$,\,$\Delta t^{\text{opt}}$), is the binomial method faster than an unbiased approach? In order to answer this question we first obtain the expected CPU time of the binomial method with the optimal choice inserting Eqs.(\ref{eq:optimal_Dt}) and (\ref{eq:number_of_realizations_optimal_dt}) in 
 Eq.~\eqref{eq:cpu_t_binomial}:
 {

\begin{equation}\label{eq:cpu_t_binomial_optimal}
 t_B^{\text{(CPU,opt)}}=\frac{27}{4} C_B \frac{|\lambda|}{\varepsilon} M\,T.
\end{equation}

 }
On the other hand, the CPU time needed to generate one trajectory using the unbiased method is proportional to the maximum time $T$, and the total CPU time to generate $M$ trajectories is {$t_U^{\text{(CPU)}}=C_UM\,T$}, where {$C_U$} is a constant depending {on} the unbiased method used. The expected ratio between the optimal CPU time consumption with the binomial method an the unbiased approach is
{

\begin{equation}\label{eq:alpha}
 \alpha=\frac{t_B^{\text{(CPU,opt)}}}{t_U^{\text{(CPU)}}}=\frac{27}{4}\frac{C_B}{C_U}\frac{|\lambda|}{\varepsilon}.
\end{equation}
}
Eq.\eqref{eq:alpha} defines what we called ``the $\frac{27}{4}$ rule", and its usefulness lies in the ability to indicate in which situations the binomial method is more efficient than the unbiased procedure (when $\alpha<1$). Also from Eq.\eqref{eq:alpha} we note that unbiased methods become the preferred option as the expected precision is increased, i.e. when $\varepsilon$ is reduced. We note that there is a threshold value {$\varepsilon_{\text{\tiny{TH}}}=\frac{27}{4}\frac{|\lambda| C_B}{C_U}$} for which both the unbiased and binomial methods are equally efficient.

Eqs.~\eqref{eq:optimal_Dt},~\eqref{eq:number_of_realizations_optimal_dt} and~\eqref{eq:alpha} conform the main result of this work. These three equations (i) fix the free parameters of the binomial method ($\Delta t$ and $M_B$) in order to compute averages with fixed precision $\varepsilon$ at minimum CPU time usage, and (ii) inform us if the binomial method is more efficient than the unbiased method. The use of these equations require the estimation of four quantities: $\sigma$, {$C_U$}, $\lambda$, and $C_B$, which can be computed numerically with limited efforts.  { While $\sigma$ and $\lambda$ rely solely on the process and approximation, hence are expected to remain constant across different machines, both {$C_U$} and $C_B$ depend on the machine, but also on the programming language and the user's ability to write efficient codes}. The standard deviation $\sigma$ depends only on the random variable $Z$ and has to be computed anyway in order to have a faithful estimate of the errors. As we will show in the examples of section~\ref{sec:numerical}, the constant $\lambda$ can be obtained through extrapolation at high values of $\Delta t$ (thus, very fast implementations). Finally, the constants {$C_U$} and $C_B$ can be determined very accurately and at a little cost by measuring the CPU usage time of a few iterations with standard clock routines. Furthermore, in Supplementary Note 2, we provide a detailed discussion on the estimation of {$C_U$} without the need to implement any {unbiased} method. This approach offers a practical means to determine the value of {$C_U$} while avoiding the complexities associated with {unbiased} methods.

{We can also work with alternative rules that fix the relative error, defined to as

\begin{equation}\label{eq:def_rel_err}
    \varepsilon_r:=\Bigg|\frac{\langle Z\rangle -Z_M}{\langle Z\rangle}\Bigg|,
\end{equation}

instead of the absolute error $\varepsilon$. To do so, we consider that the difference $\langle Z\rangle -Z_M$ is of order $\varepsilon$ and replace $\langle Z\rangle$ by a rough estimation $Z_M$. Then, we can replace in Eqs.~\eqref{eq:optimal_Dt},~\eqref{eq:number_of_realizations_optimal_dt} and~\eqref{eq:alpha}

\begin{equation}\label{eq:rel_err}
    \varepsilon\approx \varepsilon_r |Z_M|,
\end{equation}

where we note that the errors of using Eq.~\eqref{eq:rel_err} instead of Eq.~\eqref{eq:def_rel_err} are of order $\left(\varepsilon/Z_M\right)^2$. Therefore, working with \ja{relative} errors result in implicit rules, in the sense that one has to make a rough estimation of the quantity that we aim to estimate (i.e. $\langle Z\rangle$).}

{In the analysis of errors, the number of agents $N$ plays a crucial \ja{role} due to its significant impact on the magnitude of fluctuations. For instance, when estimating average densities of individuals, \ja{and when the central limit theorem applies,} the standard \ja{error} scales as $\sigma\sim1/\sqrt{N}$~\cite{van1992stochastic}. The average time $\Delta t$ between updates in unbiased methods is expected to be inversely proportional to $N$ (see Supplementary Note 2). Therefore, we expect $C_U\sim N$. Since $\lambda$ is a difference between biased and unbiased estimations, it will have the same scaling with $N$ \ja{as} the quantity $\langle Z\rangle$ (see Supplementary Note 3). The constant $C_B$ depends crucially on the method used to sample binomial random variables, and in some cases is independent \ja{of} $N$, as discussed in Supplementary Note 4. Therefore, when estimating average densities, we anticipate $\alpha$ to decrease with increasing system size, as

\begin{equation}\label{eq:scaling_alpha}
    \alpha\sim1/N,
\end{equation}

making the use of biased methods more suitable as the system size grows.}

\subsection{Numerical study}\label{sec:numerical}
In this section, we want to compare the performance of the Gillespie algorithm (in representation of the unbiased strategies) and the binomial method {(in representation of unbiased synchronous methods)}. Also, we show the applicability of the rules derived in last section to fix the optimal values of $\Delta t$ and $M_B$, and decide whether the {biased} or {unbiased} method is faster. {We will do so in the context of the SIS model with all-to-all connections and a more complex SEIR model with  meta-population structure.}

\subsubsection{{All-to-all SIS model}}\label{sec:ATA}
We study in this section the all-to-all connectivity, where every agent is connected to all others and have the same values of the transition rates. In the particular context of the SIS process, these rates read :

\begin{align}
\label{eq:individual_rates_ATA}
 & w(0\to 1)=\beta \sum_{j=1}^{N}\frac{s_j}{N}=\beta \frac{n}{N}, \quad w(1\to 0)=\mu.
\end{align}

Where $\mu$ represents the rate at which infected individuals recover from the disease and $\beta$ is the rate at which susceptible individuals contract the disease from an infected contact. The transition rates at the macroscopic description are also easily read from the macroscopic variable itself. From Eq.~(\ref{eq:def_macroscopic_transition_rates}):

\begin{align}\label{eq:macroscopic_rates_ATA}
 & W(n\rightarrow n+1)=\beta \, \frac{n}{N} (N-n) \nonumber \\
 & W(n\rightarrow n-1)=\mu \, n.
\end{align}

The main outcome of this all-to-all setting is well known and can easily be derived from the mean-field equation for the average number of infected {individuals}~\cite{marro2005nonequilibrium},

\begin{equation}
 \frac{d\langle n(t)\rangle}{dt}=\beta \, \frac{\langle n\rangle}{N}\,  (N-\langle n\rangle)-\mu \, \langle n\rangle
\end{equation}

and indicates that for {$R_0 := \beta/\mu>1$} there is an ``active'' phase with a non-zero stable steady-sate value $\langle n\rangle_{\text{st}}=(1-\mu/\beta)N$, whereas for $R_0<1$ the stable state is the ``epidemic-free'' phase $\langle n\rangle_{\text{st}}=0$ where the number of infected individuals tends to zero with time.

In order to draw trajectories of this process with the binomial method we use Algorithm~\ref{al:BM} with a single class containing all agents, $N_\ell=N,\,n_\ell=n$. The probability to use in the binomial distributions is extracted from the individual rates of Eq.~(\ref{eq:individual_rates_ATA}):

\begin{equation}\label{eq:probability_transitions}
 P(1,\Delta t)=1-e^{-\mu\, \Delta t}, \quad P(0,\Delta t)=1-e^{-\beta \, (n/N)\, \Delta t}.
\end{equation}

We note that the probability $ P(0,\Delta t)$ in Eq.\eqref{eq:probability_transitions} that a susceptible agent experiences a transition in a time $\Delta t$ is an approximation of

\begin{equation}\label{eq:exact_transmission_prob}
  1-\exp{\left(-\frac{\beta}{N} \int_t^{t+\Delta t} n(t') \,  dt'\right)}.
\end{equation}

Such approximation is a good representation of the original process when $\Delta t$ is so small that $n(t)$ can be considered as constant {in $\left[t,t+\Delta t\right]$}. In any case, we checked both analytically (see Supplementary Note 3) and numerically [see Fig.~\ref{fig:ATA_dt}-(a) and (b)] that the errors of the method still scale linearly with the time discretization, as pointed out in section \emph{The $\frac{27}{4} rule$}.
\begin{figure*}
 \centering
\includegraphics[scale=1]{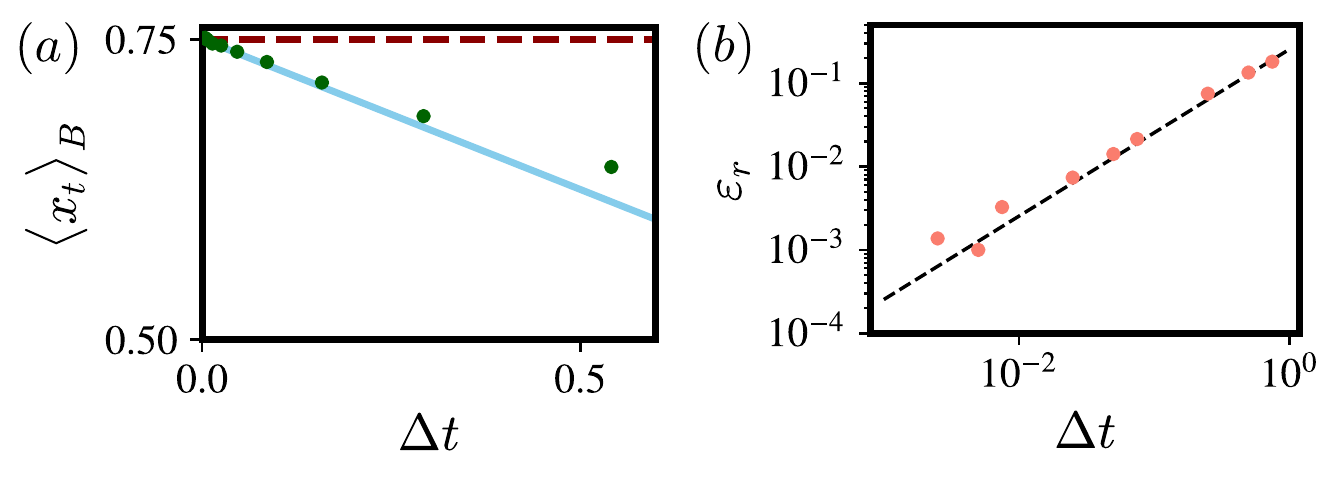}
 \caption{\textbf{Scaling of errors.} Panel (a) plots the average density $\langle x_t\rangle_B:=\dfrac{\langle n_t\rangle_B}{N}$ of infected individuals of the all-to-all SIS model at time $t=20$ obtained using the binomial method for different values of the discretization step $\Delta t$. The number of realizations is $M_B=100$, and other parameter values are $\beta=4$, $\mu=1$, $N=10^3$, $n(t=0)=10$. The statistical error bars are smaller than the symbol size. In accordance with Eq.\eqref{eq:averageZ_linear}, we find that the average $\langle x_t\rangle_B$ follows a linear dependence at small $\Delta t$ with slope $\lambda=-0.25(1)$. The horizontal dashed line is the extrapolation at $\Delta t=0$ {of $\langle x\rangle_B$} obtained from the linear fit {(continuous line)}. In panel (b) we plot for the same case, the relative error $\varepsilon_r:=\left|\dfrac{\langle n_t\rangle_B}{\langle n_t\rangle}-1\right|$, using a very accurate value of $\dfrac{\langle n_t\rangle}{N}=0.7497$ obtained with the so-called Gaussian approximation~\cite{lafuerza2010gaussian}, corroborating the linear dependence with the discretization step (dashed line of slope {$0.25\times N/\langle n_t\rangle=0.33$}).
 }
 \label{fig:ATA_dt}
\end{figure*}
 
\begin{figure*}
 \centering
\includegraphics[scale=1]{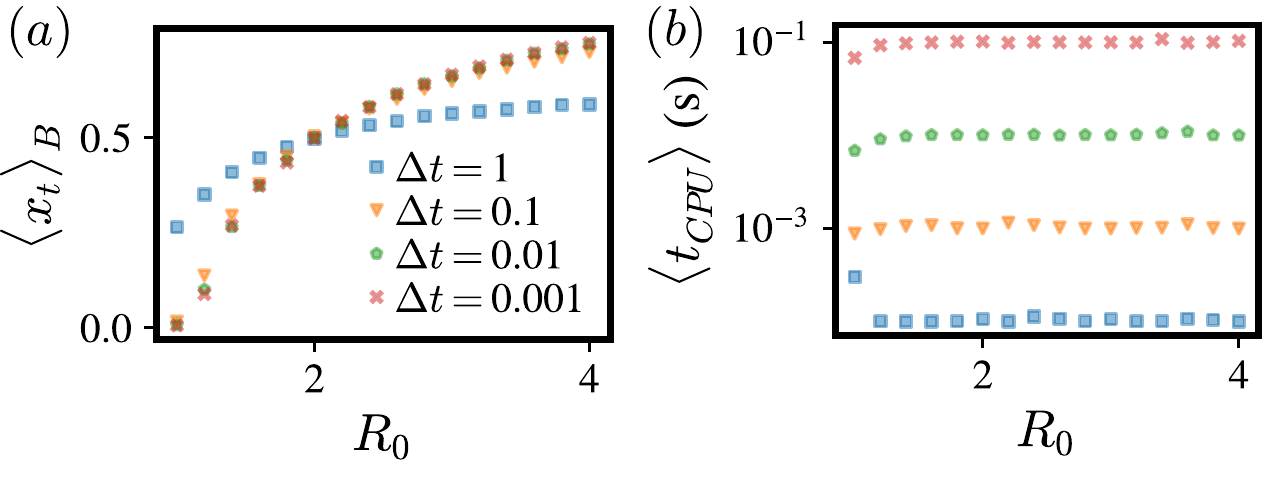}
 \caption{\textbf{Relevance of time discretization.} We plot in panel (a) the average density $\langle x_t\rangle_B:=\dfrac{\langle n_t\rangle_B}{N}$ of infected individuals of the all-to-all SIS model at time $t=20$ obtained using the binomial method as a function of $\mathcal{R}_0=\beta / \mu$ for different discretization times $\Delta t$. We take $n(t=0)=10$, $\mu=1.0$, $N=10^3$, and $M_B=100$. Statistical error bars are smaller than the symbol size. The estimations of the average agree within errors for $\Delta t=10^{-3}$ and $\Delta t=10^{-2}$. However, discrepancies are found for bigger values of $\Delta t$, for which the systematic errors are bigger than the statistical errors. Thus, the analysis of systematic errors should be taken into account to produce results with fixed desired precision . In panel (b), we plot the average CPU time (in seconds) per realization which, according to Eq.\eqref{eq:cpu_t_binomial} scales as $1/\Delta t$. This figure evidences the need of a fine tuning of $\Delta t$ in order to avoid slow and imprecise calculations. }
 \label{fig:choosing_dt}
\end{figure*}

{Now, let us illustrate the relevance of choosing an appropriate discretization $\Delta t$ for the binomial method.} First we look for a condition on $\Delta t$ that ensures that Eq.~\eqref{eq:exact_transmission_prob} can be properly approximated by Eq.~\eqref{eq:probability_transitions}. Since the average time between updates at the non-zero fixed point  $W( n_{\text{st}})^{-1}=[2\mu(1-\mu/\beta)N]^{-1}$, a heuristic sufficient condition to ensure proper integration is to fix $\Delta t \propto 1/N$. In Fig.~\ref{fig:choosing_dt}-(a), it is shown that this sufficient condition indeed generates a precise integration of the process. Also in Fig.~\ref{fig:choosing_dt}-(a) we can see that this is in contrast with the use of $\Delta t=1$, which provides a poor representation of the process (as claimed in~\cite{fennell2016limitations}). However, regarding the CPU-time consumption, the sufficient option performs poorly [Fig.~\ref{fig:choosing_dt}-(b)]. Therefore, a proper balance between precision and CPU time consumption requires to fine tune the parameter $\Delta t$. This situation highlights the relevance of the rule derived in section \emph{The $\frac{27}{4} rule$} to choose $\Delta t$ and discern if the binomial method is advantageous with respect to the unbiased counterparts.

\begin{figure*}
 \centering
\includegraphics[scale=1]{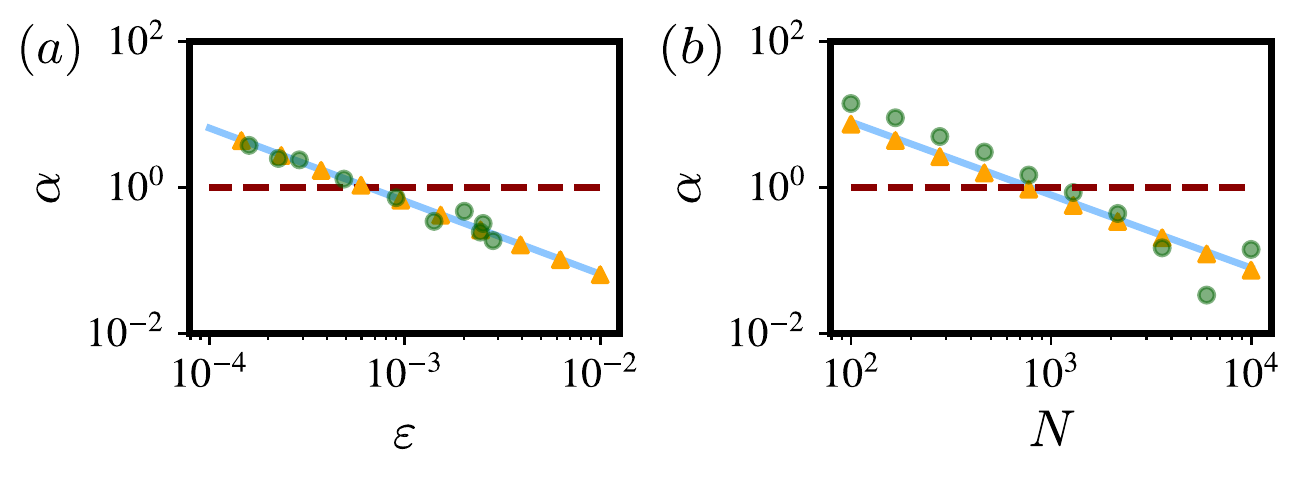}
 \caption{\textbf{27/4 rule in all-to-all SIS model.} We plot in panel (a) the ratio between the CPU times of the binomial and the Gillespie algorithms applied to the simulation of an all-to-all SIS model with parameter values $T=20$, $\mu=1$, {$\beta=4$}, $N=10^3$, and $n(t=0)=10$ as a function of the target error $\varepsilon$. The dots are the results of the numerical simulations using the binomial method with the optimal values of the discretization step $\Delta t^\text{opt}$ and number of realizations $M_B^\text{opt}$ as given by Eqs.~\eqref{eq:optimal_Dt} and~\eqref{eq:number_of_realizations_optimal_dt}, while the number of trajectories in the Gillespie algorithm was computed from Eq.~\eqref{eq:def_precision}. The solid line is Eq.~\eqref{eq:alpha}, using the values obtained from the simulations: {$\lambda=-0.25$, $C_U=7\cdot10^{-3}$ s, $C_B=2\cdot 10^{-6}$ s}. {With triangles we represent} results from the use of Eq.~\eqref{eq:alpha} with the estimation of {$C_U$} explained in Supplementary Note 2. The dashed horizontal line at $\alpha=1$ signals where the {unbiased and biased} methods are equally efficient and it crosses the data at {$\varepsilon_{_{TH}}=\frac{27}{4}\frac{|\lambda|C_B}{C_U}=3\cdot 10^{-3}$}. {In panel (b) we proceed similar to (a), but fix the precision, and {vary $N$}. Again, we fix $\Delta t$ and $M_B$ to their optimal values using Eqs.~\eqref{eq:optimal_Dt} and~\eqref{eq:number_of_realizations_optimal_dt} respectively, and plot results from simulations (dots), our prediction from Eq.~\eqref{eq:alpha} measuring $\lambda$, {$C_U$}, and $C_B$ from simulations (solid line), and Eq.~\eqref{eq:alpha} using our theoretical estimation of {$C_U$} (triangles) using Eq.~B4. This plot is in agreement with the expected scaling of $\alpha$ from Eq.~\eqref{eq:scaling_alpha}. See values of absolute CPU time consumption in Supplementary Note 5.} }
 \label{fig:ATA_rule}
\end{figure*}

In Fig.~\ref{fig:ATA_rule}-(a), we show the agreement of {Eqs.~\eqref{eq:alpha} and~\eqref{eq:scaling_alpha}} with results from simulations. In this figure, the discretization step $\Delta t$ and number of realizations for the binomial method $M_B$ have been optimally chosen according to Eqs.~\eqref{eq:optimal_Dt} and~\eqref{eq:number_of_realizations_optimal_dt}. This figure informs us that the binomial method is more efficient than an unbiased Gillespie algorithm counterpart {for a system of size $N=10^3$} when the target error is large, namely for {$\varepsilon > 3\cdot 10^{-3}$}, whereas the unbiased method should be the preferred choice for dealing with high precision estimators. { In Fig.~\ref{fig:ATA_rule}-(b) we fix the precision and vary the system size $N$ to check that $\alpha$ is inversely proportional to $N$ [Eq.~\eqref{eq:scaling_alpha}]. Thus, the efficiency of biased methods tends to overcome unbiased approaches as the system size grows. Both in Fig.~\ref{fig:ATA_rule}-(a) and (b), we show that it is possible to use estimations of {$C_U$} {without actually having to implement the unbiased method} (see Supplementary Note 2). This finding highlights the possibility of achieving accurate results while avoiding the complexities associated with implementing biased methods. It is relevant for the application of the $\frac{27}{4}$ rule that CPU time consumption is not highly dependent on $\mathcal{R}_0$ (as demonstrated in Fig.~\ref{fig:ATA_dt}-(b)). Therefore, the efficiency study can be conducted at fixed $\mathcal{R}_0$ values.}

\subsubsection{{Meta-population SEIR model}}\label{sec:meta}

{Next, we show that our results hold in a more complex model involving meta-population connectivity and many-state agents.} The meta-population framework consist on $\mathcal{C}$ sub-systems 
{or classes, such that class $\ell=1,\dots,\mathcal{C}$ contains a population of $N_\ell$ individuals}. Agents of different sub-populations are not connected and therefore cannot interact, whereas agents within the same population interact through an all-to-all scheme {similar to the one used in} Sec. \emph{All-to-all SIS model}. Individuals can diffuse through populations, thus infected individuals can move to foreign populations and susceptible individuals can contract the disease abroad. Diffusion is tuned by a mobility matrix $\textbf{m}$, being the element $m_{\ell,\ell'}$ the rate at which individuals from population $\ell$ travel to population $\ell'$. Therefore, to fully specify the state of agent $i$ we need to give its state {$s_i$} and the sub-population $\ell_i$ it belongs to at a given time. Regarding the macroscopic description of the system, the inhabitants of a population can fluctuate and therefore it is needed to keep track of all the numbers $N_\ell$ as well as the occupation numbers $n_\ell$.

{In this case we examine the SEIR paradigmatic epidemic model where agents can exist in one of four possible states: susceptible, exposed, infected, or recovered (see e.g.~\cite{keeling2011modeling}). The exposed and recovered compartments are new additions compared to the SIS model discussed in the previous section. These compartments represent individuals who have been exposed to the disease but are not yet infectious, and individuals who are immune to the disease respectively.}
The rates of all processes at the sub-population level are:
{
\begin{align}\label{eq:macroscopic_rates_meta}
    & W_\ell(S_\ell,E_\ell \rightarrow S_\ell-1,E_\ell+1)=\beta I_\ell S_\ell/N_\ell, \nonumber \\
    & W_\ell(I_\ell,E_\ell \rightarrow I_\ell+1,E_\ell+-1)=\gamma E_\ell,\nonumber \\
    & W_\ell(I_\ell,R_\ell \rightarrow I_\ell-1,R_\ell+1)=\mu I_\ell,\nonumber \\
    & W(N_\ell,N_{\ell'}\rightarrow N_\ell-1,N_{\ell'}+1)= m_{\ell,\ell'} N_\ell,
   \end{align}
}
{   
where $S_\ell$, $E_\ell$, $I_\ell$, and $R_\ell$ denote the number of susceptible, exposed, infected, and recovered individuals in population $\ell$, respectively.}

If we assume homogeneous diffusion, the elements of the mobility matrix are $m_{\ell,\ell'}=m$ if there is a connection between subpopulations $\ell$ and $\ell'$ and $m_{\ell,\ell'}=0$ otherwise. Also if the initial population distribution is homogeneous, $N_\ell(t=0)=N_0,\,\forall \ell$, then the {total} exit rate reads:
{

\begin{equation}
 W(\boldsymbol{s})=\sum_{\ell=1}^\mathcal{C} \left(\beta \, I_\ell\, \frac{S_\ell}{N_\ell}+ \gamma \,  E_\ell+\mu \,  I_\ell\right) + m \, \mathcal{C} \, N_0,
\end{equation}
}
which can be expressed as a function of the occupation variables {$\{S_\ell,E_\ell,I_\ell,N_\ell\}$}.
In this case, the average time between mobility-events, $[m\mathcal{C}N_0]^{-1}$, is constant and inversely proportional to the total number of agents $\mathcal{C} N_0$. This makes simulating meta-population models with unbiased methods computationally expensive, as a significant portion of CPU time is devoted to simulating mobility events. The binomial method is, therefore, the preferred strategy to deal with this kind of process (see Supplementary Note 6 for details on how to apply the binomial method to meta-population models~\cite{Binomials}). However, one has to bear in mind that the proper use of the binomial method requires supervising the proper value of $\Delta t$ that generates a faithful description of the process at affordable times.

\begin{figure}
 \centering
\includegraphics[scale=1]{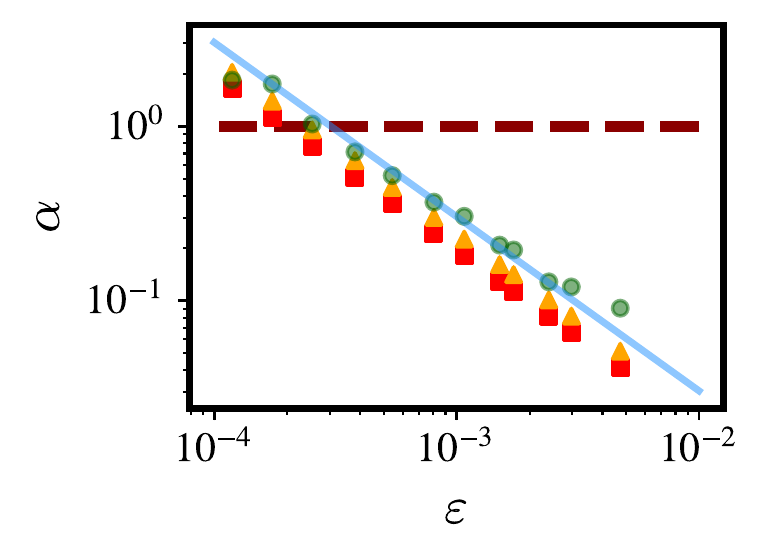}
\caption{\textbf{27/4 rule in meta-population SIS model.} Similar to Fig.\ref{fig:ATA_rule} for the case of the meta-population {SEIR} model with parameter values {$t=7.5$}, {$\gamma=1$}, $\mu=1$, {$\beta=4$}. There are ${\cal C}=100$ subpopulations arranged in a square $10\times 10$ lattice such that each subpopulation is connected to $4$ nearest neighbors (we assume periodic boundary conditions); each subpopulation contains initially $N_\ell(t=0)=10^3$ agents, $\forall \ell$. {At time zero the state of the system is $I_1(0)=10$, $I_\ell(0)=0$ $\forall \ell\ne 1$, ${E}\ell(0)=0$, $R_\ell(0)=0$, $S_\ell(0)=N_0-I_\ell$ $\forall \ell$.} We have set the mobility {among} neighboring subpopulations to a constant value $m=10$. The discretization step and the number of trajectories of the binomial method take the optimal values of Eqs.~\eqref{eq:optimal_Dt} and~\eqref{eq:number_of_realizations_optimal_dt}, while the number of trajectories in the Gillespie algorithm was computed from Eq.~\eqref{eq:def_precision}. {The required constants measured from the simulations are $\lambda=0.045$, $C_U=0.12$ s, $C_B=1.2\cdot 10^{-4}$ s}. The dashed horizontal line at $\alpha=1$ signals where the Gillespie and binomial methods are equally efficient and it crosses the data at {$\varepsilon_{_{TH}}=\frac{27}{4}\frac{|\lambda|C_B}{C_U}\approx 3\cdot 10^{-4}$}. The continuous line is the theoretical prediction Eq.\eqref{eq:alpha}, while {circles are results from simulations}. {Squares and triangles are estimations of $\alpha$ that avoid making simulations of the original process. Squares where obtained through simulations of the all-to-all process. Triangles also use the all-to-all process plus the estimation of {$C_U$} using deterministic mean-field equations as outlined in Supplementary Note 2. We note that the values of $\alpha$ are in general agreement across theory, simulations and approximations. See fit for $\lambda$ in Fig. S5 of the Supplementary Note 5.}
\label{fig:Meta_rule}}
\end{figure}

In Fig.~\ref{fig:Meta_rule} we also check the applicability of the rules derived in section \emph{The $\frac{27}{4} rule$}, this time in the context of metapopulation models. As in the case of all-to-all interactions, the preferential use of the binomial method is conditioned to the desired precision for the estimator. Indeed, unbiased methods become more convenient as the target errors decrease.

{\subsubsection{Efficient calculation of $C_U$ and $C_B$}}
{In principle, the use of Eq.~\eqref{eq:alpha} requires the implementation of both the unbiased and biased methods to estimate the constants {$C_U$} and $C_B$. It would be preferable to devise rules that do not require both implementations, as they can become cumbersome for complex processes with numerous reactions. To address this issue, we propose two approximations to Eq.\eqref{eq:alpha}. The first approximation consists of conducting the efficiency analysis on a simpler all-to-all system rather than on the meta-population structure, as outlined in Supplementary Note 2. Our second proposal entirely avoids the implementation of the unbiased method, opting instead for the mean-field estimation of $C_U$ as also described in Supplementary Note 2. In Fig.\ref{fig:Meta_rule}, we also illustrate the concurrence between these two approximations and the direct application of Eq.\eqref{eq:alpha}.}
{Overall, Fig.~\ref{fig:Meta_rule}} shows the advantage of using the binomial method for low precision. Compared to the case of the all-to-all interactions of section \emph{All-to-all SIS model}, the required CPU-time of the Gillespie method is very large, making it computationally very expensive to use. Therefore, this situation exemplifies the superiority of the binomial method with optimal choices for the discretization times and number of realizations {when dealing with complex processes.}

{\subsubsection{Final implementation}}
{Summing up, we propose the following steps to use the results of section \emph{The $\frac{27}{4} rule$}.
}
{
\begin{algorithm}[H]
    \caption{Rule 27/4}
    \label{al:27/4}
    \begin{algorithmic}[1]
    \item {Estimate a target quantity, $\langle Z\rangle$, using the biased method with several (large) values of $\Delta t$. Plotting the estimations versus $\Delta t$, compute $\lambda$ as the slope of the linear fit [see Fig.~\ref{fig:ATA_dt}-(a)  and Fig. S6 of the Supplementary Note 5 for examples].}
    \item {Estimate {$C_U$} and $C_B$ on simple all-to-all process. Alternatively, estimate {$C_U$} using deterministic mean-field calculations as in Supplementary Note 2. Estimations can be done at a small system size $N_s$, then {$C_U$} at target system size $N$ is recovered through $C_U(N)=C_U (N_s){N}/{N_s}$. }
    \item {Use Eq.~\eqref{eq:alpha} to discern whether the unbiased (for $\alpha>1$) or biased (for $\alpha<1$) approach are the most efficient option.}
    \item {If the biased method is the preferred option, then use Eqs.~\eqref{eq:optimal_Dt} and~\eqref{eq:number_of_realizations_optimal_dt} to fix the discretization time and number of realizations respectively.}
    \end{algorithmic}
   \end{algorithm}
}
\section{Conclusion}\label{sec:conclussion}

This work provides {useful insight into} the existing debate regarding the use of the binomial approximation to sample stochastic trajectories. The discretization time of the binomial method needs to be chosen carefully since large values can result in errors beyond the desired precision, while low values can produce extremely inefficient simulations. A proper balance between precision and CPU time consumption is necessary to fully exploit the potential of this approximation and make it useful.

We have demonstrated, through both numerical and analytical evidence, that the systematic errors of the binomial method scale linearly with the discretization time. Using this result, we can establish a rule for selecting the optimal discretization time and number of simulations required to estimate averages with a fixed precision while minimizing CPU time consumption. {Furthermore, when comparing specific biased and unbiased implementations, we have derived a rule to identify the more efficient option.}

{ It is not possible to determine whether the unbiased or biased approach is the best option in absolute terms. CPU time consumption varies depending on factors such as the programming language, the machine used for calculations, and the user's coding proficiency. This variability is parametrized through the constants {$C_U$} and $C_B$ in our theory. Nevertheless, we can make general statements independent of the implementation. Firstly, } the advantage of using the binomial method depends on the target precision: the use of unbiased methods becomes more optimal as the target precision increases. {Second, since CPU time scaling with the number of reactions depends on the method, biased methods tend to outperform unbiased methods as the complexity of the model increases.}

The numerical study of our proposed rules signals that the ratio of CPU times between the unbiased and binomial methods are similar in both all-to-all and meta-population structures. This result facilitates the use of the rules in the latter case. Indeed, one can develop the study of efficiency in the all-to-all framework and then use the optimal values of the discretization time and number of realizations in the more complex case of meta-populations. 

{Our work contributes to the generation of trustworthy and fast stochastic simulations, crucial for many real-world applications. Future work will focus on generalizing this approach to \ja{deal with adaptive discretizations} and address cases involving non-Poissonian processes (see e.g.~\cite{ferguson2006strategies}), where unbiased algorithms are challenging to implement. }

\section*{Code availability}

The codes for the different models are available at~\cite{Github_J_Aguilar} and are free to use
providing the right credit to the author is given.

{Different binomial random number generators from different authors were used for robustness analysis~\cite{Binomials_netlib,Binomials,source_binomial,kachitvichyanukul1988binomial,press1996numerical,DAVIS1993205,fishman1979sampling}.  Check Supplementary Note 4 for details.}

\section*{Data availability}
Data sharing not applicable to this article as no datasets were generated or analysed during the current study.

\section*{Author contribution}

J.A., J.J.R., and R.T. conceived and designed the study. J.A. performed the simulations. J.A., J.J.R., and R.T. wrote the paper. All the authors read and approved the paper.

\section*{Competing interests}
The authors declare no competing interests.

\begin{acknowledgments}
We thank Sandro Meloni and~{Luis Irisarri} for useful {comments}. \ja{We also thank anonymous reviewers for valuable comments that have helped us to improve the presentation and discussion of our results.} Partial financial support has been received from the Agencia Estatal de Investigaci\'on (AEI, MCI, Spain) MCIN/AEI/10.13039/501100011033 and Fondo Europeo de Desarrollo Regional (FEDER, UE) under Project APASOS (PID2021-122256NB-C21 and PID2021-122256NB-C22) and the María de Maeztu Program for units of Excellence in R\&D, grant CEX2021-001164-M, and the Conselleria d'Educaci\'o, Universitat i Recerca of the Balearic Islands (grant FPI\_006\_2020), {and the contract ForInDoc (GOIB)}. 

\end{acknowledgments}

\bibliography{refs}

\clearpage
\newpage
\onecolumngrid
 \clearpage

\setcounter{page}{1}
\setcounter{figure}{0}
\setcounter{equation}{0}
\setcounter{section}{0}

\renewcommand{\thefigure}{SI \arabic{figure}}

\renewcommand{\thesection}{Supplementary Note \arabic{section}} 
\renewcommand{\theequation}{S\arabic{equation}}
\renewcommand{\thefigure}{S\arabic{figure}}
\setcounter{equation}{0}

\begin{center}
 \Large{SUPPLEMENTARY INFORMATION\\ Biased versus unbiased numerical methods for stochastic simulations: \\
 application to contagion processes\\ }
\end{center}
\begin{center}
 Javier Aguilar, Jos\'e J. Ramasco, and Ra\'ul Toral. \\
Instituto de F\'{\i}sica Interdisciplinar y Sistemas Complejos IFISC (CSIC-UIB), Campus UIB, 07122 Palma de Mallorca, Spain. \\
Departamento de Electromagnetismo y Física de la Materia e Instituto Carlos I de Física Teórica y Computacional, Universidad de Granada, Granada E-18071, Spain . \\
\end{center}
\hrulefill

\section{Optimal time}\label{AP:optimal_time}
In this section, we derive expressions equivalent to Eqs.~\eqref{eq:optimal_Dt},~\eqref{eq:number_of_realizations_optimal_dt}, and ~\eqref{eq:alpha} in the main text but for arbitrary scaling relations. First, we consider the situation in which the  systematic errors of the biased method scale as an arbitrary power of the discretization time,
\begin{equation}
    \langle Z \rangle = Z_{M_B}+\lambda \left(\Delta t\right)^\varphi \pm \varepsilon_B.
\end{equation}
Considering this generalization is pertinent since sub-linear and super-linear scalings ($\varphi<1$ and $\varphi>1$ respectively) can occur within the context of numerical methods for simulating stochastic models, for example, in the context of the numerical integration of stochastic differential equations~\cite{toral2014stochastic,asmussen2007stochastic}. We also consider the case in which the scaling of the CPU time with the number of realizations can be sublinear,
\begin{equation}\label{eq:cpu_t_binomial_V2}
 t_B^{\text{(CPU)}}=C_B \left(M_B\right)^\psi \frac{ T}{\Delta t},
\end{equation}
and
\begin{equation}\label{eq:cpu_t_unbiased_V2}
 t_U^{\text{(CPU)}}=C_U M^\psi T.
\end{equation}
Non-trivial values for $\psi$ could be obtained when working on massively parallel architectures like GPUs.

Since error sources can only accumulate, the only way to reduce errors in biased computations is to increase the number of realizations ($M_B$). The errors using a biased method equals the ones of the unbiased counterpart when
\begin{equation}\label{eq:number_of_realizations_V2} M_B=M\left(\frac{|\lambda|\left(\Delta t\right)^\varphi}{\varepsilon}-1\right)^{-2}.
\end{equation}
Inserting Eq.~\eqref{eq:number_of_realizations_V2} in Eq.~\eqref{eq:cpu_t_binomial_V2} we obtain:

\begin{equation}\label{eq:t_CPU_B_vs_delta_t}
 t_B^{\text{(CPU)}}=\frac{M C_B T}{\Delta t}\left(\frac{|\lambda|\left(\Delta t\right)^\varphi}{\varepsilon}-1\right)^{-2\psi}.
\end{equation}

The above equation informs about the CPU time consumption using the binomial method with a time discretization $\Delta t$ with general scaling relations. Eq.\eqref{eq:t_CPU_B_vs_delta_t} has two relative extrema. We discard one of them, $(\Delta t)^\varphi = \varepsilon/|\lambda|$, since its use would require sampling infinite-many biased trajectories [see Eq.~\eqref{eq:number_of_realizations_V2}]. The other relative extrema is a minimum that we identify as the optimal value for $\Delta t$:

\begin{equation}\label{eq:optimal_Dt_V2}
 \Delta t ^ \text{opt} = \left[\frac{1}{1+2\varphi\psi} \frac{\varepsilon}{|\lambda|}\right]^{\frac{1}{\varphi}}.
\end{equation}

Inserting this result in Eq.~\eqref{eq:number_of_realizations_V2} we obtain the general expression for the optimal number of realizations to be sampled with the biased method
\begin{equation}\label{eq:number_of_realizations_optimal_dt_V2}
    M_B = M\left(1+\frac{1}{2\varphi\psi}\right)^2.
\end{equation}
Lastly, the general ratio of CPU times reads
\begin{equation}\label{eq:alpha_V2}
 \alpha=\frac{t_B^{\text{(CPU,opt)}}}{t_U^{\text{(CPU)}}}=\left(2\varphi\psi\right)^{-2\psi}\left(1+2\varphi\psi\right)^{2\psi+\frac{1}{\varphi}}\frac{C_B}{C_U}\left(\frac{|\lambda|}{\varepsilon}\right)^{\frac{1}{\varphi}}.
\end{equation}
Setting $\psi=\varphi=1$, Eqs.~\eqref{eq:optimal_Dt_V2},~\eqref{eq:number_of_realizations_optimal_dt_V2}, and~\eqref{eq:alpha_V2} reduce to Eqs.~\eqref{eq:optimal_Dt},~\eqref{eq:number_of_realizations_optimal_dt}, and~\eqref{eq:alpha} respectively.
\\
\section{Estimation and scaling of the constant C}\label{AP:estimation_C}
In the main text, we use the constant {$C_U$} to characterize the CPU time usage of unbiased methods,

\begin{equation}\label{eq:T_CPU_U}
t_U^{\text{(CPU)}} = C_U M\,T.
\end{equation}

Where $M$ is the number of realizations and $T$ is the simulation time at which the program stops. The constant {$C_U$} is employed in Eq.\eqref{eq:alpha} to determine the relative efficiency between the biased and unbiased methods for computing averages with fixed precision. While it is possible to measure {$C_U$} with limited computational efforts, it still requires the implementation of an unbiased algorithm. In this section, we propose estimations of {$C_U$} that circumvent such implementation, simplifying the use of Eq.\eqref{eq:alpha}.

\subsection{{Deterministic mean-field approximation}}\label{AP:deterministic_MF}
{For the first approximation, we}  rewrite Eq.~\eqref{eq:T_CPU_U} in terms of the average CPU time to execute one iteration with the unbiased method ($t_{U,\text{it}}^{\text{(CPU)}}$) and the average number of iterations required to simulate one trajectory of duration $T$ ($\mathcal{N}_T$),

\begin{equation}\label{eq:estimation_CPU_t}
    t_U^{\text{(CPU)}} = t_{U,\text{it}}^{\text{(CPU)}} \mathcal{N}_T M.
\end{equation}

To proceed with the theoretical estimation of  {$\mathcal{N}_T$}, we rewrite it as

\begin{equation}
    \mathcal{N}_T=\frac{T}{\langle \Delta t\rangle_T},
\end{equation}

where $\langle \Delta t\rangle_T$ is the average time between updates of the process along a trajectory of duration $T$.  {Hence, we can estimate the constant {$C_U$} as

\begin{equation}\label{eq:estimation_CPU_t_V2}
    C_U\approx\frac{t_{U,\text{it}}^{\text{(CPU)}}}{\langle \Delta t\rangle_T}.
\end{equation}

}
{
The advantage of using Eq.~\eqref{eq:estimation_CPU_t_V2} comes from the fact that we can estimate $t_{U,\text{it}}^{\text{(CPU)}}$ without implementing the unbiased method in a program. Instead, we can measure it as the average CPU time consumption that it takes to execute the operations involved in one iteration of the exact algorithm.
Furthermore, we propose to estimate $\langle \Delta t\rangle_T$ using the deterministic mean field approximation of the process,
}

\begin{equation}\label{eq:mean_time_T}
    \langle \Delta t\rangle_T=\frac{1}{T}\int_0^T\frac{1}{W\left[ n(t) \right]}dt.
\end{equation}

\begin{figure*}[h!]
    \centering
   \includegraphics[scale=0.9]{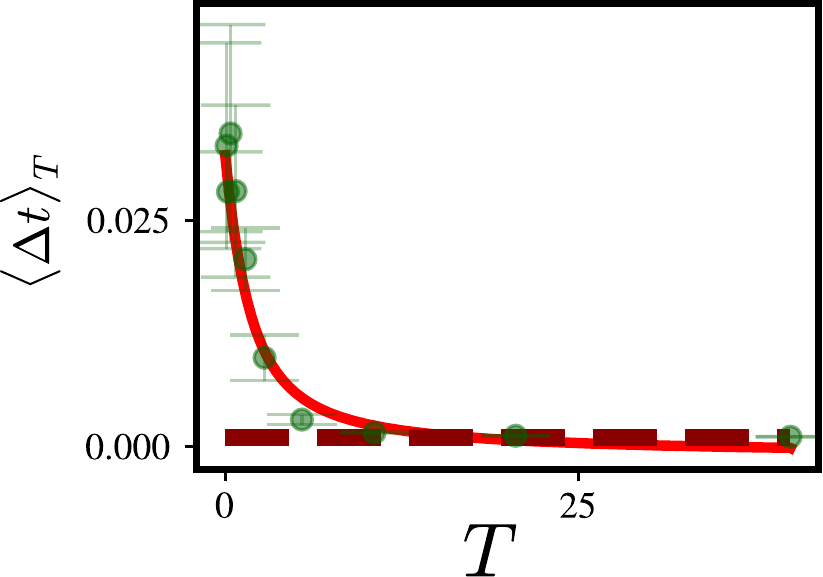}
   \caption{{Estimation of $\langle \Delta t\rangle_T$ for different values of the final time $T$ using the mean-field formula from Eq.~\eqref{eq:estimation_dt} (solid line), its long-time limit (Eq.~\eqref{eq:long_time_limit_dt} in dashed line), and measures from simulations (dots and errorbars noting the $97.5^\text{th}$ porcentile). Parameters: $N=1000$, $\beta=2$, $\mu=1$.}
   }\label{AP_fig:estimation_of_dt}
\end{figure*}

Where $W\left[ n(t) \right]$ is the total exit rate defined in Eq.~\eqref{eq:exit_rate} evaluated in the state of the system at time $t$, and using the deterministic mean-field dynamics. {For example, in the case of the all-to-all SIS model,

\begin{equation}
    W\left[ n(t) \right]=n(t)\left( \beta \frac{N-n(t)}{N}+\mu\right),
\end{equation}

where $n(t)=N x(t)$, and $x(t)$ is the solution to 

\begin{equation}
    \frac{d}{dt}x(t)=x(t)\left( \beta (1-x(t))-\mu\right),
\end{equation}

which is

\begin{equation}
    x(t)=\frac{e^{(\beta-\mu) t}}{\frac{R_0 \left(e^{(\beta-\mu)
    t}-1\right)}{R_0-1}+\frac{1}{x_0}},
\end{equation}

The integral in Eq.~\ref{eq:mean_time_T} for the SIS model reads
%
%
\begin{eqnarray}
    \langle \Delta t\rangle_T&=&\frac{R_0 }{2\mu (R_0 -1)}-\frac{\left(1-e^{(1-R_0 ) \mu T}\right) (R_0  (x_{_0}-1)+1)}{(R_0 -1)^2 (R_0 +1) \mu T x_{_0}}\nonumber \\
    &&-\frac{R_0  }{2 (R_0 +1)^2 \mu T}\log \left[\frac{(R_0 -1)
    (R_0  (x_{_0}-1)-1)}{(R_0 +1) e^{T-R_0  \mu T} (R_0  (x_{_0}-1)+1)-2 R_0  x_{_0}}\right],\label{eq:estimation_dt}
\end{eqnarray}
with

\begin{equation}
    R_0=\frac{\beta}{\mu}.
\end{equation}

}
{For the particular case of systems with stable or meta-stable states, like the SIS all-to-all model, one can further approximate this derivation and evaluate Eq.~\eqref{eq:mean_time_T} in the non-zero stable state ($n_\text{st}$). 

\begin{equation}\label{eq:long_time_limit_dt}
   \lim_{T\to\infty} \langle \Delta t\rangle_T=\frac{1}{W[n_\text{st}]}=\frac{R_0 }{2\mu (R_0 -1)}.
\end{equation}

In Fig.~\ref{AP_fig:estimation_of_dt}, we show the agreement between Eqs.~\eqref{eq:estimation_dt} and~\eqref{eq:long_time_limit_dt} and results from simulations.
}

{
In order to produce the dotted line in Fig.~\ref{fig:ATA_rule}-(a), we used Eqs.~\eqref{eq:estimation_CPU_t_V2} and~\eqref{eq:estimation_dt}, where we measured $ t_{U,\text{it}}^{\text{(CPU)}}$ as the time needed to generate one uniform random number, make its logarithm, execute an \emph{if-} statement, and carry the arithmetic operations that it would require to compute the exit rate.}

\subsubsection{{Scaling with system size}}
{
As we can see through this example, it is generally true that the total exit rate is extensive in the number of agents, this is, we can express the exit rate as

\begin{equation}
    W\left[ n(t) \right]=N f\left[ x(t) \right],
\end{equation}

where $f\left[ x(t) \right]$ is an intensive function. Putting all together, the CPU time with the unbiased method reads

\begin{equation}\label{eq:estimation_CPU_t_V1}
    t_U^{\text{(CPU)}}=\frac{t_{U,\text{it}}^{\text{(CPU)}}  T N}{\int_0^T f^{-1}(t)dt}  M T,
\end{equation}

where we can identify an estimation of {$C_U$},

\begin{equation}\label{eq:estimation_C_V1}
    C_U= \frac{t_{U,\text{it}}^{\text{(CPU)}}  T N}{\int_0^T f^{-1}(t)dt}.
\end{equation}

Therefore, we expect that {$C_U$} scales, at least, linearly with the system size. Depending on the process, the scaling could be super-linear, since  $t_{U,\text{it}}^{\text{(CPU)}}$ depends on the number of reactions, which could depend itself on the number of agents. Thus, it is possible to do the efficiency study on small systems (of size $N_s<<N$) and then scale to the values of $C_U$ at big $N$ with}
{

\begin{equation}
    C_U(N)=C_U (N_s){N}/{N_s}
\end{equation}

}

\subsubsection{{SEIR model}}

{Eq.~\eqref{eq:estimation_CPU_t_V2} can also be used to approximate the constant {$C_U$} for multidimensional models, like the SEIR model of section \emph{Meta-population SEIR model}. The deterministic mean-field equations of the SEIR model read

\begin{equation}\label{eq:MF_SEIR}
    \begin{cases}
    \frac{d}{dt}s(t)=-\beta s(t) i(t),\\
    \frac{d}{dt}e(t)=\beta s(t) i(t)-\gamma e(t),\\
    \frac{d}{dt}i(t)=\gamma e(t)-\mu i(t),\\
    \frac{d}{dt}r(t)=\mu i(t).\\
    \end{cases}
\end{equation}

With $s:=S/N$, $e:=E/N$, $i:=I/N$, $r:=R/N$, and $N=S+E+I+R$. One more time, we can estimate the average time between updates with the integral 

\begin{equation}\label{eq:dt_jEIR}
    \langle \Delta t\rangle_T=\frac{1}{T}\int_0^T\frac{1}{W\left[ \boldsymbol{s}(t) \right]}dt.
\end{equation}

Where the total exit rate for the all-to-all SEIR model reads

\begin{equation}
    W\left[ \boldsymbol{s}(t) \right]=W\left[ S,E,I,R \right]=\beta I\frac{S}{N}+\gamma E+\mu I.
\end{equation}

The triangles in Fig.~\ref{fig:Meta_rule} where obtained integrating Eqs.~\eqref{eq:MF_SEIR} and~\eqref{eq:dt_jEIR} numerically with $N=1000$, $\beta=4$, $\gamma=1$,$\mu=1$, $S(t=0)=N-10$, $I(t=0)=10$, $E(t=0)=R(t=0)$, $T=2.4$.
}

\subsection{{All-to-all approximation}}\label{AP:ata_approx}
{
The second approach that we propose can be applied to meta-population models. The idea is to assess efficiency,  as outlined in section \emph{The $\frac{27}{4} rule$}, but based on a simplified all-to-all model with $N$ agents. Thus ignoring the meta-population structure. For the case of the SEIR model described in section \emph{Meta-population SEIR model}, the rates for its associated all-to-all version read,}
{
\begin{align}
    & W(S,E \rightarrow S-1,E+1)=\beta \,I\, S/N, \nonumber \\
    & W(I,E \rightarrow I+1,E+-1)=\gamma \,E,\nonumber \\
    & W(I,R \rightarrow I-1,R+1)=\mu \,I,\nonumber \\
    & N=S+E+I+R.
\end{align}
}
{Let $C_U^\text{ATA}$ and $C_B^\text{ATA}$ be the constants appearing in  Eq.~\eqref{eq:alpha} for the all-to-all setting, then we approximate the constants {$C_U$} and $C_B$ associated to the meta-population system as }
{
\begin{equation}
    C_B\approx C_B^\text{ATA} \quad C_U\approx C_U^\text{ATA} \frac{N}{N_0\,\mathcal{C_U}},
\end{equation}
}
{where we assume the scalings with the system size for {$C_U$} and $C_B$ discussed respectively in Supplementary Note 2, and~\ref{AP:estimation_CB}. In this way, one substitutes the implementations on a meta-population system with $\mathcal{C_U}\cdot N_0$ agents with much simpler all-to-all implementations with $N$ agents.}

\section{{Estimation and scaling with system size of  the constant $\lambda$}}\label{AP:errors}

Consider a SIS model with all-to-all interactions, and let $n(t)$ be the number of infected individuals at time $t$. The probability that a susceptible agent will change its state at time {$t'\in[t,t+\Delta t]$} is:

\begin{equation}\label{AP-eq:probability_to_change_state_real}
 P(0,\Delta t)_\text{exact}=1-\exp{\left(-\frac{\beta}{N} \int_t^{t+\Delta t} n(s) ds\right)}.
\end{equation}

In the context of the binomial approximation, this probability is approximated by:

\begin{equation}\label{AP-eq:probability_to_change_state_approx}
 P(0,\Delta t)=1-\exp{\left(-\frac{\beta}{N} n(t) \Delta t\right)}.
\end{equation}

The difference between Eqs.~(\ref{AP-eq:probability_to_change_state_real}) and (\ref{AP-eq:probability_to_change_state_approx}) is the error associated to the use of Eq.~(\ref{AP-eq:probability_to_change_state_approx}) instead of Eq.(\ref{AP-eq:probability_to_change_state_real}). We call this difference $\Delta P$.

\begin{equation}\label{AP-eq:Difference_in_P}
 \Delta P= P(0,\Delta t)_\text{exact}-P(0,\Delta t)=\exp{\left(-\frac{\beta}{N} n(t) \Delta t\right)}-\exp{\left(-\frac{\beta}{N} \int_t^{t+\Delta t} n(s) ds\right)}.
\end{equation}

Considering $\Delta t$ small, we can approximate
 
\begin{equation}
 \int_t^{t+\Delta t} n(s) ds\approx n(t)\Delta t+ \dot{n}(t)\frac{\Delta t^2}{2}.
 \end{equation}

 {Where $\dot{n}(t)=\frac{d}{dt}n(t)$. Inserting the above expression in Eq.~(\ref{AP-eq:Difference_in_P}), we obtain
 \begin{eqnarray}\label{AP-eq:Difference_in_P_V2}
 \Delta P&=& \exp{\left(-\frac{\beta}{N} n(t) \Delta t\right)}-\exp{\left(-\frac{\beta}{N} \left[n(t)\Delta t+ \dot{n}(t)\frac{\Delta t^2}{2}\right]\right)}\nonumber\\
 &=&\exp{\left(-\frac{\beta}{N} n(t) \Delta t\right)}\left[1-\exp{\left(-\frac{\beta}{N} \dot{n}(t)\frac{\Delta t^2}{2}\right)}\right]\approx \frac{\beta \dot{n}(t)}{2N}\Delta t^2.
\end{eqnarray}
 }
If we make use of the binomial method, the faithful increment in the number of infected individuals should be $\Delta n_\text{exact}$, a random variable drawn from a binomial distribution ${\bf B}\left(n(t),P(0,\Delta t)_\text{exact}\right)$. Instead, we use a random variable $\Delta n$ drawn from the approximate distribution ${\bf B}\left(n(t),P(0,\Delta t)\right)$. The difference between the mean values of the exact random variable and the actual one used in the numerical method is 

\begin{equation}
 \langle\Delta n\rangle_\text{exact} - \langle\Delta n\rangle= n(t)\Delta P\sim \Delta t^2.
\end{equation}

If we aim to reach a final simulation time $T$, the accumulated error of using the approximation Eq.~(\ref{AP-eq:Difference_in_P}) for a number of iterations proportional to $T/\Delta t$ scales as $\Delta P /\Delta t\sim \Delta t$. This scaling is corroborated numerically in Fig.~\ref{fig:ATA_dt} of the main text.

{Therefore, the estimation for $\lambda$ associated to the number of infected individuals scales linearly with $N$,

\begin{equation}
    \lambda \sim \mathcal{O}(N),
\end{equation}

whereas the errors over densities [$x(t)$] are independent of $N$,

\begin{equation}
    \lambda \sim \mathcal{O}(1),
\end{equation}

}

\section{{Scaling of the constant $C_B$}}\label{AP:estimation_CB}
{
The scaling of the constant $C_B$ with the parameters of the process depends crucially on the algorithm used to generate binomial samples, i.e. random numbers ${\bold B}(N,p)$ from a binomial distribution [Eq.\eqref{eq:binomial}] of parameters $N$ and $p$, representing the number of trials and the probability of success, respectively. There are several algorithms to generate ${\bold B}(N,p)$ numbers and which one is more convenient (i.e. faster, using less CPU execution time) depends on the specific values of the parameters $N$ and $p$ and, to a lesser extent, on the machine and programming language used. Generally speaking, the methods split in two families: those whose CPU time grows linearly on $N$ and those whose CPU time is independent of $N$. Amongst the first, we mention the numerical inversion of the cumulative distribution function, and the iteration of $N$ Bernouilli processes each one with success probability $p$. Amongst the second set of methods, the simplest one is a rejection algorithm in which a value selected from a Lorenztian distribution is accepted with a carefully chosen probability. A more sophisticated inversion method whose execution time is also independent of $N$ is discussed in Ref.~\cite{kachitvichyanukul1988binomial}. Other methods whose execution time is independent of the system size exploit the relation between binomial and beta distributions (see e.g. ~\cite{fishman1979sampling}).
}

{
One tries to profit from the best of each family by choosing methods whose time scale linearly with $N$ up to a threshold number of trials in which it is more profitable to switch to an $N$-independent algorithm for sufficiently large $N$. In general, the criterion when to choose from one method to another is determined by a threshold value $\Theta_\text{TH}$ of the product $Np$ (the expected value of the binomial random variable). Furthermore, as a significant fraction of the time needed for the binomial method of Algorithm~\ref{al:BM} is spent in the generation of the random numbers, the above strategy translates into an approximately linear dependence of the time $C_B$, which depends on the discretization time $\Delta t$ through Eq.~\eqref{eq:prob_remaining_constant} and on the number of individuals $N$ through the dependence of the occupation numbers $n_\ell$, up to the threshold and constant afterwards,
}
{

\begin{equation}
    C_B(N,\Delta t)=
    \begin{cases}
        c_1 N \quad \text{if } Np(N,\Delta t)<\Theta_{\text{TH}}.\\
        c_2 \quad \text{otherwise },
    \end{cases}
\end{equation}

where $c_1$ and $c_2$ are constant numbers depending on the generation method used.
}

{
We have considered three different random number generator for the binomial distribution:\\
(i) The first one is the built-in Python function {\tt numpy.random.binomial}~\cite{source_binomial} that uses the method based on the inverse of the cumulative function up to   $\Theta_{\text{TH}}=30$, and the sophisticated algorithm discussed in~\cite{kachitvichyanukul1988binomial} otherwise. We have used this generator in all the simulations of the paper.  \\
(ii) The second is the Fortran routine {\tt ZBQLBIN}  available as part of the  {\tt randgen} package at ~\cite{Binomials}. This method is based on sampling from a beta distribution which can be related to the target binomial~\cite{fishman1979sampling}.\\
(iii) Based on the rejection method implemented in the Fortran routine {\tt BNLDEV} of Ref.~\cite{press1996numerical} and our own timing tests we have written a Fortran routine {\tt iran\_bin} which uses the following scheme~\cite{Github_J_Aguilar}:
\begin{itemize}
\item If $p<0.15$ and $pN<15$ use inversion of the cumulative distribution.
\item if $p>0.15$ and $N<100$  use repetition of Bernouilli processes.
\item In all other cases use the routine {\tt BNLDEV} based on rejection.
\end{itemize}
}

\begin{figure*}[h!]
    \centering
   \includegraphics[scale=0.85]{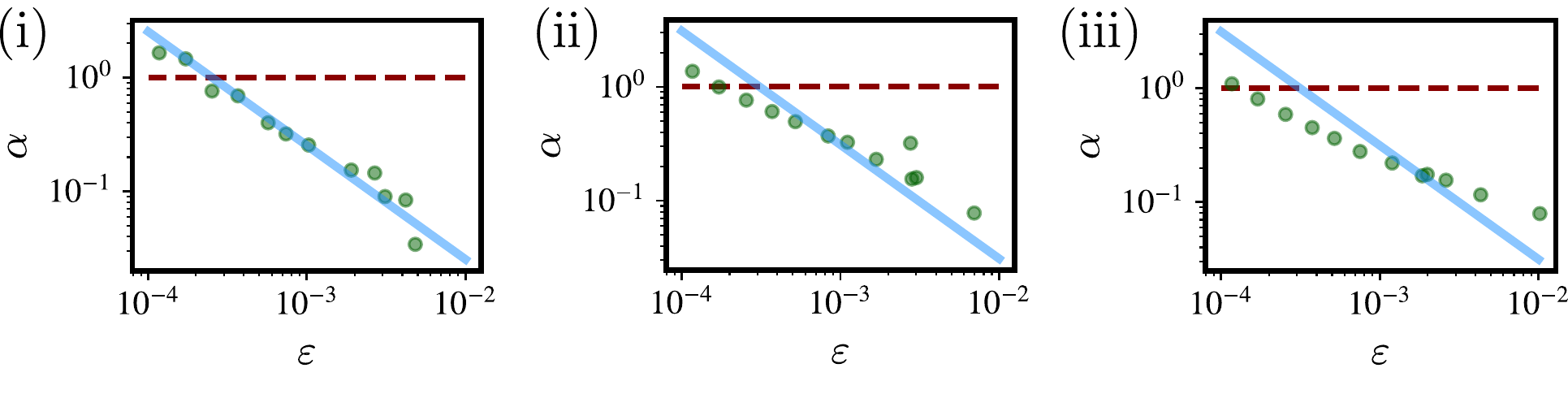}
   \caption{{The figure shows the values of $\alpha$ [Eq.~\eqref{eq:alpha}] for different target precisions $\varepsilon$, and the different algorithms explained in the text to extract binomial samples (Methods i), ii), and iii)). The process is the metapopulation SEIR model described in section  \emph{Meta-population SEIR model} with the same parameters as used in Fig.~\ref{fig:Meta_rule}. Dots are the results from simulation while the continuous line is the prediction of our theory.}
   }\label{AP_fig:alpha_SEIR}
\end{figure*}

\begin{figure*}[h!]
    \centering
   \includegraphics[scale=0.85]{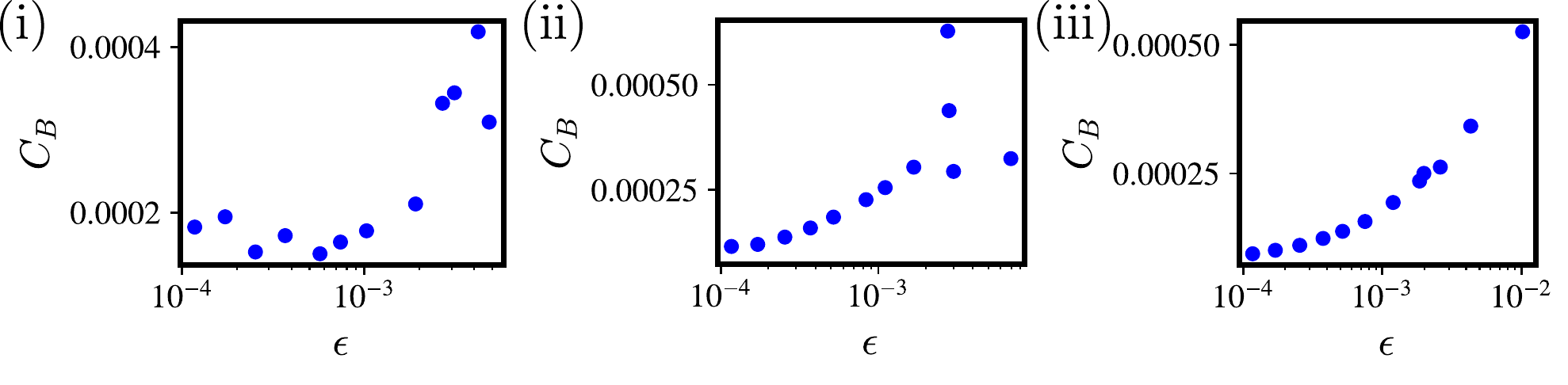}
   \caption{ {The figure shows the values of $C_B$ measured from simulations corresponding to dots in Fig.~\ref{AP_fig:alpha_SEIR}.}}\label{AP_fig:CB_SEIR}
\end{figure*}

{
In Fig.~\ref{AP_fig:alpha_SEIR} we plot the value of $\alpha$ defined in Eq.~\eqref{eq:cpu_t_binomial_optimal} for each of the three generators applied to the metapopulation SEIR model described in section \emph{Meta-population SEIR model} with the same parameters as used in Fig.~\ref{fig:Meta_rule}. Defining the threshold error value $\epsilon_\text{TH}$ as the minimum value for which the binomial method is more efficient than the Gillespie algorithm, i.e. the one for which $\alpha=1$, we find that $\epsilon_\text{TH}[\text{generator (iii)}]<\epsilon_\text{TH}\left[\text{generator (ii)}\right]<\epsilon_\text{TH}\left[\text{generator (ii)}\right]$, showing that generator (iii) is the most efficient one. In the same figure, we also plot the theoretical line associated to Eq.~\eqref{eq:alpha}, which, overall offers a good approximation to the numerical values. Some discrepancies can be attributed to the dependence of $C_B$ on the error $\varepsilon$. To clarify this point, we plot in Fig.~\ref{AP_fig:CB_SEIR} the values of $C_B$ estimated from the same simulations used in Fig.~\ref{AP_fig:alpha_SEIR} as

\begin{equation}
    C_B=\frac{\Delta t^\text{opt}}{T}\frac{t_B^{(\text{CPU})}}{M^\text{opt}_B},
\end{equation}

where $t_B^{(\text{CPU})}$ is the time needed to do the biased simulations used in Fig.~\ref{AP_fig:alpha_SEIR}.
}

\begin{figure*}[h!]
    \centering
   \includegraphics[scale=0.85]{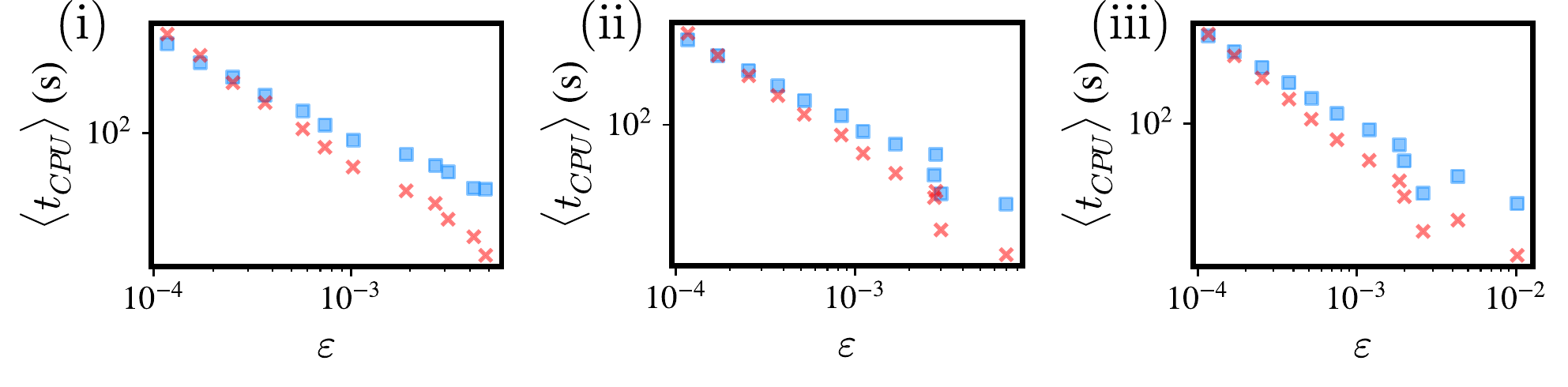}
   \caption{ {The figure shows the absolute times measured from simulations corresponding to dots in Fig.~\ref{AP_fig:alpha_SEIR} for both the unbiased method (squares) and the biased method (x).}}\label{AP_fig:absolute_times_SEIR}
\end{figure*}

{
In Fig.~\ref{AP_fig:absolute_times_SEIR} we show the absolute times used to generate dots in Fig.~\ref{AP_fig:alpha_SEIR} with both the biased and unbiased methods. This figure illustrates that all methods operate within similar temporal scales.
}

%
%

\section{{Auxiliary figures}}
\subsection{{Absolute CPU time consumption}}\label{AP:absolute_times}
{
    In this section, we present the absolute time counterparts of Fig.~\ref{fig:ATA_rule} instead of its ratio, denoted as $\alpha$.
}
\begin{figure*}[h!]
    \centering
   \includegraphics[scale=1]{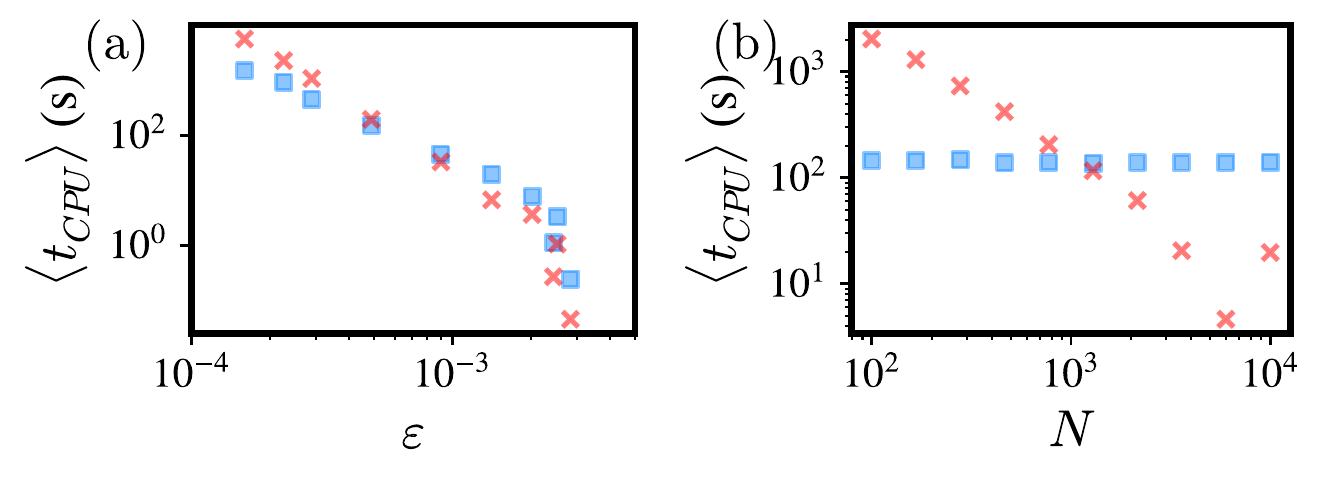}
    \caption{{Average CPU time consumption using the binomial and Gillespie methods (x and squares respectively). The tasks for $(a)$ and $(b)$ correspond to those of Figs.~\ref{fig:ATA_rule} (a) and (b) respectively.}} 
    \label{fig:metapop_SEIR_errors}
\end{figure*}

\subsection{{Scaling of errors SEIR}}

\begin{figure*}[h!]
    \centering
   \includegraphics[scale=1]{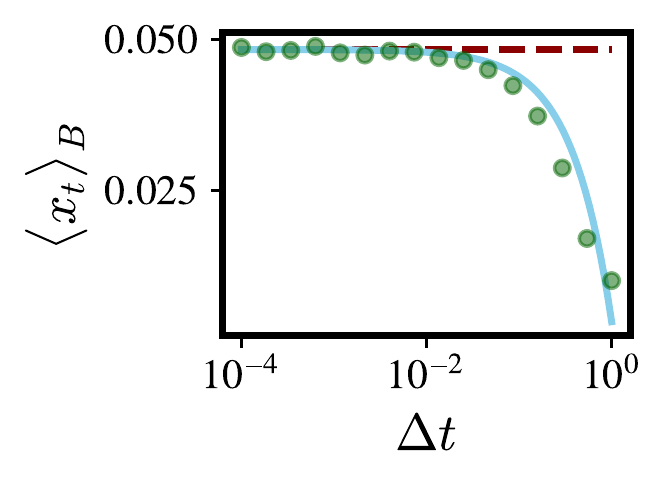}
    \caption{{Average density $\langle x_t\rangle_B:=\dfrac{\langle I_t\rangle_B}{N_0 M}$ of infected individuals of the metapopulation SEIR model at time $t=7.5$ obtained using the binomial method for different values of the discretization step $\Delta t$. The number of realizations is $M_B=100$, and other parameter values are the same of Fig.~\ref{fig:Meta_rule}.  Circles are the results from simulations and the continuous line is a linear fit whose slope is $\lambda=-0.045(1)$. The horizontal dashed line is the extrapolation at $\Delta t=0$ {of $\langle x\rangle_B$} obtained from the linear fit.}} 
    \label{fig:ATA_absolute_times_metapop}
\end{figure*}

\clearpage

\section{Binomial method on meta-population framework}\label{AP:Binomial_metapopulation}

{
In this section, we show how to adapt the binomial method (algorithm~\ref{al:BM}) to the case of meta-population models with SEIR dynamics (described in section \emph{Meta-population SEIR model}). Let $S_\ell(t)$, $E_\ell(t)$, $I_\ell(t)$, and $R_\ell(t)$ be, respectively, the number of susceptible, exposed, infected, and recovered individuals in subpopulation $\ell=1,\dots,\mathcal{C}$ at time $t$. These occupation numbers fully characterize the state of the system. Note that the total number of agents in class $\ell$ at time $t$ is $N_\ell(t)=S_\ell(t)+E_\ell(t)+I_\ell(t)+R_\ell(t)$. We partition mobility and epidemic events and perform separate updates for each of them to sample the future state $\{S_\ell(t+\Delta t),E_\ell(t+\Delta t),I_\ell(t+\Delta t),R_\ell(t+\Delta t)\}_{\ell=1,\dots,\mathcal{C}}$.
}

{
\emph{-Mobility: } The first step involves the calculation, for all sub-populations, of the number of agents who move within a time interval $\Delta t$. These quantities, denoted by $\{\Delta X_{\ell}\}_{\ell=1,\dots,\mathcal{C}}$, for $X=S,E,I,R$, are extracted from binomial distributions $\Delta X_\ell\sim {\bf B}\left(X_\ell(t),p^\text{out}_\ell\right)$,  with $p^\text{out}_\ell:=1-e^{-\Delta t\sum_{j}m_{\ell,j}}$. Then, traveling agents have to be distributed among neighboring sub-populations. We call $\Delta X_{\ell,\ell'}$, respectively, the number of agents from compartment $X$ entering in sub-population $\ell'$ coming from $\ell$. Those numbers are sampled from the multinomial distributions, ${\bf M}(\Delta X_\ell;{\{p_{\ell,\ell'}\}_{\ell'=1,\dots,\mathcal{C}}})$  with $p_{\ell,\ell'}:=\dfrac{m_{\ell,\ell'}}{\sum_{j}m_{\ell,j}}$.} The general multinomial distribution ${\bf M}(N;p_1,\dots,p_k)$ is defined by the probabilities 

\begin{equation}
 P(n_1,\dots,n_k)={N\choose {n_1\cdots n_k}}p_1^{n_{1}}\dots p_k^{n_{k}}.
\end{equation}

One possible method for sampling numbers $\{n_1,\dots,n_k\}$ from a multinomial distribution is by using an ordered sequence of binomial samples~\cite{DAVIS1993205}.

\begin{equation}
 n_i\sim {\bf B}\left(N-\sum_{j<i}n_j,\frac{p_{i}}{1-\sum_{j<i}p_j}\right),\quad i=1,\dots,k.
\end{equation}

At this point, the state of the system is updated with the mobility events:
\begin{eqnarray}\label{eq:mobility_update_n}
 X_\ell(t)&\leftarrow& X_\ell(t)+\sum_j \Delta X_{j,\ell},\label{eq:mobility_update_s}
\end{eqnarray}
but time is not yet increased, as the changes due to epidemic dynamics still need to be accounted for.\\

{
\emph{-Epidemics:} Once agents have been reallocated according to the mobility dynamics [Eq.~(\ref{eq:mobility_update_n})], occupation numbers are updated following the epidemic rules in [Eq.~\eqref{eq:macroscopic_rates_meta}]. To do so, we extract the binomial numbers:

\begin{align}
& \Delta {n}_{\ell,S\to E} \sim{\bf B}\left[S_\ell(t),1-\exp\left({-\beta \frac{I_\ell(t)}{N_\ell(t)}\Delta t}\right)\right], \nonumber \\
& \Delta n_{\ell,E\to I}\sim {\bf B}\left(E_\ell(t),1-e^{-\gamma \Delta t}\right), \nonumber \\
&\Delta n_{\ell,I\to R}\sim {\bf B}\left(I_\ell(t),1-e^{-\mu \Delta t}\right),
\end{align}

The new state of the system reads,

\begin{align}
 & S_\ell(t+\Delta t)=S_\ell(t)-\Delta {n}_{\ell,S\to E}, \nonumber \\ 
 & E_\ell(t+\Delta t)=E_\ell(t)+\Delta {n}_{\ell,S\to E}- \Delta {n}_{\ell,E\to I}, \nonumber \\ 
 & I_\ell(t+\Delta t)=I_\ell(t)+\Delta {n}_{\ell,E\to I}- \Delta {n}_{\ell,I\to R}, \nonumber \\ 
 & R_\ell(t+\Delta t)=R_\ell(t)+\Delta {n}_{\ell,I\to R}.
\end{align}

Finally, time is updated $t\to t+\Delta t$.
}

\end{document}